\documentclass[aps,prc,nofootinbib,showpacs,superscriptaddress,groupedaddress]{revtex4-1}
\usepackage{amsmath,amssymb,amsbsy,bm}
\usepackage{graphicx}
\usepackage{comment}
\usepackage{float}
\usepackage[colorlinks=true,linkcolor=blue,citecolor=blue,urlcolor=blue]{hyperref}

\begin{document}

\title{Charge Conservation and Higher Moments of Charge Fluctuations}
\author{Scott Pratt}
\affiliation{Department of Physics and Astronomy and National Superconducting Cyclotron Laboratory\\
Michigan State University, East Lansing, MI 48824~~USA}
\author{Rachel Steinhorst}
\affiliation{Department of Physics and Astronomy\\
Michigan State University, East Lansing, MI 48824~~USA}
\date{\today}


\begin{abstract}
Higher moments of distributions of net charge and baryon number in heavy-ion collisions have been proposed as signals of fundamental QCD phase transitions. In order to better understand background processes for these observables, models are presented which enable one to gauge the effects of local charge conservation, decays of resonances and clusters, Bose symmetrization, and volume fluctuations. Monte Carlo methods for generating samplings of particles consistent with local charge conservation are presented, and are followed by a review of simple analytic models involving a single type of charge with a constant experimental efficiency. The main model consists of thermal emission superimposed onto a simple parameterization of collective flow, known as a blast-wave, with emission being consistent with individual canonical ensembles. The spatial extent of local charge conservation is parameterized by the size and extent over which charge is conserved. The sensitivity of third and fourth order moments, skewness and kurtosis, to these parameters, and to beam energy and baryon density is explored. Comparisons with STAR data show that a significant part of the observed non-Poissonian fluctuations in net-proton fluctuations are explained by charge and baryon-number conservation, but that measurements of the STAR collaboration for fluctuations of net electric charge significantly differ from expectations of the models presented here.
\end{abstract}

\maketitle

\section{Introduction}\label{sec:intro}

The fluctuation of conserved charges is a standard means by which to investigate and classify phase transitions. At the critical point correlation lengths diverge, which results in peaks in charge fluctuations as one approaches the critical point. For systems with first-order phase transitions, fluctuations turn into phase separation and fluctuation measures are no longer intensive quantities. The growth of fluctuations becomes increasingly dramatic as one considers progressively higher-order fluctuations. In a volume $V$, fluctuations of a charge $Q$ can be defined as
\begin{eqnarray}\label{eq:kappadef}
\mathcal{M}_N&\equiv&\frac{1}{V}\langle(Q-\overline{Q})^N\rangle=\frac{1}{V}\sum_n P_n(n-\overline{n})^N,
\end{eqnarray}
when particles have unit charge. The measure is increasingly sensitive to the tails of the multiplicity distribution, $P_n$, as $n$ increases. The free energy, $F\sim a_2(n-\overline{n})^2 + a_3(n-\overline{n})^3\cdots$, is minimized for $n=\overline{n}$, but the quadratic part vanishes at the critical point, $a_2\rightarrow 0$, which allows the fluctuations to grow and be dominated by higher-order terms. The shapes of the tails of the distribution are then profoundly altered. 

The properties of the QCD  transition, deconfinement and the restoration of chiral symmetry, are not well understood at finite baryon density. There exists the possibility that this transition is a true phase transition, with a critical point at several times nuclear density and with a critical temperature close to the pion mass. If this is the case, it begs the question as to whether the conditions for phase separation or for critical phenomena can be reproduced in the laboratory. Heavy ion collisions at high energy, measured at the Relativistic Heavy Ion Collider (RHIC) or at the LHC, can produce mesoscopic regions at temperatures of a few hundred MeV, which is well above the expectations for a critical temperature, and densities of several times nuclear matter density. These densities might or might not be sufficiently high to investigate the phase transition. 

High-energy heavy ion collisions are characterized by strong explosive collective flow. Measurement is confined to the outgoing asymptotic momenta, but because of strong flow, correlations in coordinate space manifest themselves as correlations in relative momentum. Thus, measurements of correlations binned by relative momentum or charge fluctuations within some defined region of momentum space serve as surrogates for the corresponding observables in coordinate space. Indeed, measurements of charge and baryon number fluctuations have been performed at RHIC. By adjusting the beam energy of the colliding nuclei, experiments at RHIC have explored conditions at which novel phase phenomena might occur. Fluctuations of electric charge and baryon number have been especially popular. An initial scan of beam energies \cite{Abelev:2008jg,Sarkar:2014fja,Xu:2016hxf,Thader:2016gpa,Luo:2012kja,Tarnowsky:2011vk,Aggarwal:2010wy,Chatterjee:2019msq,Adam:2019xmk,Adamczyk:2014fia,Sarkar:2013qla,Luo:2013saa,Sahoo:2012pfx,Luo:2011ts,Sahoo:2011at,R.Sahoo:2011qtk,Nayak:2009wc,McDonald:2012ts,McDonald:2013aoa,Adam:2020unf,Adamczyk:2017wsl} was rather inconclusive, but measurements with greatly improved statistics are currently being undertaken and analyzed. The main thrust of these studies is to search for evidence of a QCD phase transition with a critical point at finite baryon density, with the hope that the phenomena one expects for an idealized equilibrated system \cite{Stephanov:1998dy,Stephanov:1999zu} becomes manifest in the measured debris of heavy-ion collisions. Even if no true phase transition exists, charge fluctuations, or equivalently susceptibilities, are fundamental properties of the quark gluon plasma and can be investigated with lattice gauge theory \cite{Borsanyi:2011sw}.

In addition to the finite system size (event multiplicities might number in the thousands), the novel states of matter created in heavy-ion collisions persist for $\lesssim 10$ fm/$c$. This severely limits the degree to which phases can separate or to which critical fluctuations can grow. This also limits the extent to which conserved charges can separate from one another. For example, if a strange and an anti-strange quanta are produced together their separation is limited by diffusion, which is difficult to calculate in lattice gauge theory \cite{Aarts:2014nba,Amato:2013naa}, but can be roughly extracted experimentally and is gaining attention theoretically \cite{Aziz:2004qu,Kapusta:2014dja,Kapusta:2017hfi,Hammelmann:2018ath}. To justify thermal models of fluctuations of conserved charge based on equilibration, sufficient time is required for particles to enter and exit some defining volume.

The first goal of this paper is to gauge the degree to which charge-balance correlations affect higher-order correlations. Charge balance functions, which are two-particle correlations related to charge conservation, have been measured extensively \cite{Wang:2012jua,Abelev:2010ab,Li:2011zzx,Alt:2007hk,Adams:2003kg,Aggarwal:2010ya,Alt:2004gx,Abelev:2013csa,Adamczyk:2015yga,Adamczyk:2013hsi,Abelev:2009ac}, and modeled theoretically \cite{Bass:2000az,Bozek:2004dt,Bozek:2003qi,Pratt:2015jsa,Pratt:2017oyf,Ling:2013ksb,Cheng:2004zy,Pratt:2012dz}. In addition to making it difficult for phases to separate or for critical correlations to emerge, local charge conservation also represents its own source of correlation, which needs to be understood as a potential source of background before making firm arguments to have observed phenomena related to phase transitions. It is well known that charge-balance correlations are readily measurable at the two particle level, $N=2$ in Eq. (\ref{eq:kappadef}), and that they explain the bulk of the $N=2$ fluctuation measurement. However, their impact on $N=3,4$ fluctuations has not be investigated in great detail. For instance, the four-particle measure of correlation,
\begin{eqnarray}
C_4&=&\mathcal{M}_4-3\mathcal{M}_2^2,
\end{eqnarray}
which is based on cumulants, subtracts much of the contribution to $\mathcal{M}_4$ coming from purely two-particle correlations. However, charge conservation can involve multiple particles, and the degree to which a cumulant-based measure, like the kurtosis, is affected by charge conservation is not fully understood. Relations based on a uniform acceptance probability and for a single type of charge were worked out in \cite{Savchuk:2019xfg}, and provide significant insight into how higher-order correlations are affected by charge conservation. The goal of this study is to extend such ideas to a more realistic picture, which takes into account the conservation of all three types of charge (baryon number, electric charge and strangeness) and applies a more realistic model of experimental acceptance and efficiency. The interplay of charge conservation with chemical equilibrium, decays, and Bose statistics are all considered.

To understand the role of chemical equilibrium and decays, a model is presented which creates small volumes in which the net charges $B,Q$ and $S$ are each fixed at some value. Even if the net charges are all zero, charged particles exist in combinations that conserve the net charge. Theoretical methods for exact calculation of the canonical ensemble and a method for Monte Carlo generation of statistically independent events are presented here. The method lends itself to including decays and accounting for experimental acceptance and efficiency. The physical picture of treating small volumes as independent patches was previously done for calculation of charge balance functions in \cite{Schlichting:2010qia,Schlichting:2010na,Pratt:2010zn}, and was also applied in \cite{Oliinychenko:2020cmr}. Following the terminology in \cite{Oliinychenko:2020cmr}, we sometimes refer to these sub-volumes as patches. The method presented here creates perfectly independent samplings, and can generate billions of such patches within a few hours. This enables highly accurate calculations of higher moments with minimal numeric cost.

After a brief review of cumulants and the definitions of skewness and kurtosis in Sec. \ref{sec:cumulants}, the method for exact calculation of the canonical ensemble describing a multi-component, multi-charge hadron gas is presented in Sec. \ref{sec:theoryexact}. These techniques extend those used for canonical ensembles used to study isospin fluctuations of a hadron gas \cite{Cheng:2002jb}, nuclear fragmentation \cite{Pratt:1999ht}, the level density of a Fermi gas \cite{Pratt:1999ns}, and the effect of restricting a quark-gluon plasma (QGP) to having fixed charge, including being in an overall color singlet \cite{Pratt:2003jd}. Exact methods for calculating correlations up to fourth-order are presented. Unfortunately, when including realistic acceptance effects and complex decays, the exact expressions are no longer viable. However, as shown in Sec. \ref{sec:theoryMC}, the exact expressions show how sample events can be generated. Each event, defined by a set of particles and momenta, is generated with perfect independence from the others, and perfectly reproduces the canonical expressions from Sec. \ref{sec:theoryexact}. Section \ref{sec:bose} extends the previous sections to show how Bose correlations can be included. Section \ref{sec:uniformeff} considers the case of a single type of charge and a uniform efficiency, i.e. the probability of any particle being recorded is set to some fixed value. Much of this discussion repeats what is said in \cite{Savchuk:2019xfg}, and is included for completeness. This provides for a physical discussion of how charge conservation, volume fluctuations, chemical equilibrium, decays, clustering, and Bose corrections should affect higher moments.

The heart of this study is presented in Sec. \ref{sec:blast}. Here, the patches are assigned collective velocities consistent with the collective flow deduced from heavy-ion collisions. A canonical sampling of particles is generated from each patch, followed by a simulation of their decays. The particles are then overlaid onto the acceptance of the STAR detector at RHIC. Each patch is uncorrelated with any other patch, so moments can be calculated by averaging over the independent contributions from the patches. Results are displayed alongside results from the STAR Collaboration. The size and sign of the fluctuations of the net-proton distributions are consistent with observations, but the calculations of the net charge distributions differs qualitatively from STAR observations. A detailed discussion of the lessons derived from this study is presented in Sec. \ref{sec:summary}.
% !TEX root =  CCmoments.tex

\section{Cumulants, Skewness and Kurtosis}\label{sec:cumulants}

For the manuscript to be more self-contained, a brief review of cumulants and the definition of skewness and kurtosis is presented here. Cumulants of a charge distribution are defined by 
\begin{align}
C_1 &= \langle Q\rangle,\\ \nonumber
C_2 &= \langle(Q-\overline{Q})^2\rangle,\\ \nonumber
C_3 &= \langle(Q-\overline{Q})^3\rangle,\\ \nonumber
C_4 &= \langle(Q-\overline{Q})^4\rangle-C_2^2, \nonumber
\end{align}
where $Q$ is the net charge. Here, $Q$ might refer to baryon number, to strangeness, or to the electric charge measured in units of $e$. Rather than showing the cumulants $C_n$, ratios are presented to help minimize trivial dependences on system size. The skewness, $S$, is a measure of the third moment,
\begin{eqnarray}
S&=\frac{C_3}{C_2^{3/2}}.
\end{eqnarray}
This definition has the advantage in being dimensionless, but it does not become independent of volume in the limit of large volumes. Thus, it is more common to consider the ratio
\begin{eqnarray}
S\sigma&=&S\sqrt{C_2}=\frac{C_3}{C_2},
\end{eqnarray}
which becomes an intensive measure in the limit of larger volumes. However, in this study we consider the ratio
\begin{eqnarray}
\frac{S\sigma^3}{C_1}&=&\frac{C_3}{C_1},
\end{eqnarray}
which is also intensive, and approaches unity for a uncorrelated emission, i.e. a Skellam distribution.

The kurtosis is a measure of four-particle correlations,
\begin{eqnarray}
K&=&\frac{C_4}{C_2^2},
\end{eqnarray}
but instead of $K$, one typically chooses
\begin{eqnarray}
K\sigma^2&=&\frac{C_4}{C_2},
\end{eqnarray}
where $\sigma^2=C_2$, to find an intensive measure of the fluctuation. For a measure to be intensive, it should be independent of volume in the large-volume limit. For small volumes, charge conservation alters the average densities of various species, which is known as canonical suppression. Canonical suppression also distorts the higher moments for smaller volumes.

The ratios $C_4/C_2$ and $C_3/C_1$ approach simple values in the limit that the distributions would be Poissonian. For Poissonian emission the observation of a charge in one region of momentum space is uncorrelated with the emission into any other space. Thus, particles are correlated only with themselves. If charges appear only in integral positive units, one can apply the usual expression for the Poissonian moments where the mean is $\eta$, 
\begin{align}
C_1&=\overline{n}=\eta\\ \nonumber
C_2&=\langle(n-\overline{n})^2\rangle=\eta,\\ \nonumber
C_3&=\langle(n-\overline{n})^3\rangle=\eta=C_1,\\ \nonumber
C_4&=\langle(n-\overline{n})^4\rangle-3C_2^2=\eta. \nonumber
\end{align}
If there exist both positive and negative charges, the distribution of the net charge can be derived by convoluting the two distributions. Convoluting two Poissonians results in a Skellam distribution. If the mean number of positives is $\eta_+$ and the mean number of negatives is $\eta_-$, the distribution of net charge for a Skellam distribution, $Q=n_+-n_-$, yields the following cumulants
\begin{align}
\label{eq:skellam}
C_1&=\eta_+-\eta_-,\\ \nonumber
C_2&=\eta_++\eta_-,\\ \nonumber
C_3&=\eta_+-\eta_-=C_1,\\ \nonumber
C_4&=\eta_++\eta_-=C_2. \nonumber
\end{align}
Thus, if charges are produced in an uncorrelated fashion in increments of $\pm 1$, the skewness and kurtosis become
\begin{eqnarray}
S\frac{\sigma^3}{C_1}&=&\frac{C_3}{C_1}=1,\\ \nonumber
K\sigma^2&=&\frac{C_4}{\sigma^2}=1, \nonumber
\end{eqnarray}
where $\sigma^2\equiv\langle(Q-\overline{Q})^2\rangle=\eta_++\eta_-$. Even though most of the literature focuses on $S\sigma=C_3/C_2$, this study presents results for $C_3/C_1$ so that one can better understand deviations from the uncorrelated baseline. 

Moments depend on the efficiency $\alpha$ with which particles are measured. In the limit of vanishing efficiency all distributions of positives or of negatives tend to become Poissonian \cite{Bzdak:2012ab}, and the distribution of the net charge will thus become Skellam. Then Eq. (\ref{eq:skellam}) shows that as $\alpha\rightarrow 0$, $C_4/C_2=C_3/C_1=1$. This can be understood by seeing that as $\alpha\rightarrow 0$, the moments are dominated by the probability of observing either zero charges or a single charge. The probability of observing a single charge is $\alpha\rightarrow 0$, while the probability of observing two charges is proportional to $\alpha^2$, which is negligible. This assumption would fall through if multitple charges were observed on individual particle, but for final-state hadrons the charges are only $\pm 1$.

If the net charge is fixed, a non-perfect efficiency is required to produce fluctuations. For fixed charge, the efficiency divides particles into two sets, the measured and non-measured. Each set fluctuates equally, but oppositely, relative to the mean. Thus, all the even moments of net charge will have an even reflection symmetry about an efficiency of $1/2$, and the odd moments will have an odd symmetry,
\begin{eqnarray}
\label{eq:alphasymm}
C_n(1/2+\delta\alpha)&=\left\{\begin{array}{rl}
C_n(1/2-\delta\alpha),&n=2,4,6\cdots\\
-C_n(1/2-\delta\alpha),&n=3,5,7\cdots~.\end{array}\right.
\end{eqnarray}
% !TEX root =  CCmoments.tex

\section{Recursive Techniques for Generating Canonical Partition Functions}\label{sec:theoryexact}

For non-interacting particles the canonical partition function can be calculated exactly, or at least to the level that all partitions of $A\le A_{\rm max}$ hadrons are taken into account, with the exact solution being reached at $A_{\rm max}=\infty$. For our case, we conserve three quantities: the electric charge $Q$, the baryon number $B$ and the strangeness $S$. For states $i$ with energies $E_i$, the partition function,
\begin{eqnarray}
Z(Q,B,S)&=&\sum_{i,Q_i=Q,B_i=B,S_i=S}e^{-\beta E_i},
\end{eqnarray}
where $Q_i,B_i$ and $S_i$ are the discrete values of the conserved quantities for the state $i$, can be calculated recursively. The function $Z_A(Q,B,S)$ refers to the subset of states with $A$ hadrons,
\begin{eqnarray}
Z(Q,B,S)&=&\sum_{A\ge 0}Z_A(Q,B,S).
\end{eqnarray}
The recursive procedure begins with
\begin{eqnarray}
Z_{A=0}(0,0,0)&=&1,
\end{eqnarray}
the canonical partition function of the vacuum. The contribution for a given $A$, $Z_A(Q,B,S)$, can be written as 
\begin{eqnarray}\label{eq:recurrence}
Z_A(Q,B,S)&=&\frac{1}{A}\sum_h z_hZ_{A-1}(Q-q_h,B-b_h,S-s_h),
\end{eqnarray}
where $z_h$ is the single-particle partition function for hadron species $h$, which has charges $q_h,b_h$ and $s_h$. This was proved in \cite{Pratt:1999ht}, and can be understood by realizing that one can count all the ways to arrange $A$ hadrons with a given charge by considering all the ways to arrange one hadron multiplied by all the ways to arrange the remaining hadrons. To avoid double counting, a factor of $1/A$ is applied. For a fixed charge the probability to have $A$ hadrons is,
\begin{eqnarray}
P(A)&=&\frac{Z_A(Q,B,S)}{\sum_A Z_A(Q,B,S)}=\frac{Z_A(Q,B,S)}{Z(Q,B,S)}.
\end{eqnarray}
In practice, the sum over $A$ is cut off at some $A_{\rm max}$, but in our studies here that cutoff is made large enough that contributions to $Z$ for $A>A_{\rm max}$ are negligible. Thus, once one builds the partition function from $A=0$ to $A_{\rm max}$ one has the partition function for all $Q,B,S$. 

Once the partition function is calculated one can also calculate the multiplicities and moments of observing specific species. For example, the multiplicity of species $h$ in a system with charge $Q,B,S$ is
\begin{eqnarray}
\langle N_h\rangle &=& z_h\frac{Z(Q-q_h,B-b_h,S-s_h)}{Z(Q,B,S)}.
\end{eqnarray}
This also provides expressions for the various charges, e.g.,
\begin{eqnarray}
\langle Q\rangle &=& \sum_h q_hz_h\frac{Z(Q-q_h,B-b_h,S-s_h)}{Z(Q,B,S)}.
\end{eqnarray}
Spectra can also be calculated. For species $h$ with spin $j_h$,
\begin{eqnarray}
\frac{dN_h}{d^3p}&=&\frac{(2j_h+1)\Omega}{(2\pi\hbar)^3}e^{-E_h(p)}\frac{Z(Q-q_h,B-b_h,S-s_h)}{Z(Q,B,S)}.
\end{eqnarray}
Second-order moments can also be calculated exactly,
\begin{eqnarray}
\langle N_hN_{h'}\rangle &=& \delta_{hh'}z_h\frac{Z(Q-q_h,B-b_h,S-s_h)}{Z(Q,B,S)}+z_hz_{h'}\frac{Z(Q-q_h-q_{h'},B-b_h-b_{h'},S-s_h-s_{h'})}{Z(Q,B,S)}.
\end{eqnarray}
It is straightforward to extend this expression to higher-order fluctuations.

These expressions can also be extended to consider non-additive conservation laws. Net isospin conservation of a hadron gas was invoked in \cite{Cheng:2002jb}, i.e. restricting the states to being in an iso-singlet.  Quark-gluon states restricted to being in both an iso-singlet and a color singlet were addressed in \cite{Pratt:2003jd}. Bose and Fermi corrections are discussed in Sec. \ref{sec:bose}.

For the case of a single kind of charge, one can see how the the recursive method above yields the same result as what one would expect by writing down the partition function for a system with $(A-Q)/2$ negative charges and $(A+Q)/2$ positive charges as shown in \cite{Savchuk:2019xfg},
\begin{eqnarray}
\label{eq:singlecharge}
Z_{A,Q}=\left\{\begin{array}{rl}
\frac{z^A}{[(A-Q)/2]![(A+Q)/2]!},&A-Q~{\rm is~even},\\
0,&A-Q~{\rm is~odd}.\end{array}\right. ,
\end{eqnarray}
where $z$ is the partition function of a single charge. One can readily see that this is consistent with the recurrence relations,
\begin{eqnarray}
Z_{A,Q}&=&\frac{z}{A}\left\{Z_{A-1,Q-1}+Z_{A-1,Q+1}\right\}\\
\nonumber
&=&\frac{z}{A}\left\{\frac{z^{A-1}}{[(A-Q-2)/2]![(A+Q)/2]!}+\frac{z^{A-2}}{[(A-Q)/2]![(A+Q-2)/2]!}\right\}\\
\nonumber
&=&\frac{z}{A}\left\{\frac{z^{A-1}(A-Q)/2}{[(A-Q)/2]![(A+Q)/2]!}+\frac{z^{A-2}(A+Q)/2}{[(A-Q)/2]![(A+Q)/2]!}\right\}\\
&=&\frac{z^A}{[(A-Q)/2]![(A+Q)/2]!}.
\end{eqnarray}
This result is also equivalent to expectations based on setting reaction rates equal. If one assumes that pairs are created with some rate $\beta$, and that they are destroyed with some rate $\alpha N_+N_-$, where $N_+$ and $N_-$ are the number of positive and negative charges, $N_++N_-=A$. Setting the rates equal,
\begin{equation}\label{eq:ratesequal}
\alpha \frac{(A-Q)(A+Q)}{4}Z_{A,Q}=\beta Z_{A-2,Q}.
\end{equation}
One can see that if one chooses $\beta/\alpha=z$ that Eq.s (\ref{eq:ratesequal}) and (\ref{eq:singlecharge}) are consistent. If the net charge is zero, the result is even simpler,
\begin{eqnarray}
\label{eq:PAgivenQcanonical}
P(A|Q)&=&\frac{Z_{A,Q}}{\sum_A Z_{A,Q}},\\
\nonumber
P(A|Q=0)&=& \frac{z^A}{[(A/2)!]^2}\left\{\sum_{A={\rm even}} \frac{z^A}{[(A/2)!]^2}\right\}^{-1}.
\end{eqnarray}

Aside from the assumptions that $Q$ is fixed and that there exists only one kind of charge, Eq.(\ref{eq:ratesequal}) also requires that Bose and Fermi quantum statistical corrections are negligible, and that only unit charges exist. Despite these shortcomings, this picture is useful in that it allows one to see how multiplicity fluctuations are affected by charge conservation in a simple model. 
% !TEX root =  CCmoments.tex

\section{Generation of Uncorrelated Sample Events}\label{sec:theoryMC}

Complicated experimental acceptances are difficult to incorporate into expressions for the moments. It is then easiest to generate entire events via Monte Carlo, and filter the events through the acceptance. The Monte Carlo procedure involves choosing a hadron proportional to the number of ways the system might have such hadron, i.e. a product of the partition function of the individual hadron multiplied by the partition function of the remainder. The procedure becomes:
\begin{enumerate}
\item Calculate and store the partition function, $Z_A(Q,B,S)$, up to some size $A\le A_{\rm max}$ for values of  $Q,B,S$ that might ultimately couple back to a given $A=A_{\rm max}$ for the given total values $Q,B,S$. 
\item For total charge $Q,B,S$, choose the number of hadrons $A$ proportional to $Z_A(Q,B,S)/Z(Q,B,S)$, where $Z(Q,B,S) = \sum_{A\le A_{max}}Z_A(Q,B,S)$.
\item Choose a hadron $h$ proportional to the probability $z_hZ_{A-1}(Q-q_h,B-b_h,S-s_h)/Z_A(Q,B,S)$. If Bose degeneracy is to be taken into account this procedure is slightly modified as described in Sec. \ref{sec:bose}.
\item Choose the momentum proportional to the thermal weight $e^{-E_p/T}$.
\item Repeat (3,4) but with $A,Q,B,S$ being replaced by $A-1,Q-q_h,B-b_h,S-s_h$. The procedure is finished when $A=0$.
\end{enumerate}
Bose effects can be included by altering the second and third steps above. This is addressed in Sec. \ref{sec:bose}.

Storing the partition function can require substantial memory for large $A_{\rm max}$ because the indices $Q,B$ and $S$ must also vary over a range of order $\pm A_{\rm max}$, so memory usage roughly scales with $A_{\rm max}^4$. Because one is usually interested in calculations with total charge near zero, one can ignore partition functions for charges that cannot couple back to the fixed overall charge at $A_{\rm max}$. Once $A$ exceeds $A_{\rm max}/2$, the calculations here cut off values of $Q,B$ and $S$ that could not ultimately affect the $Q=B=S=0$ partition function for $A=A_{\rm max}$. Even with this savings, partition functions with $A_{\rm max}=250$ could require approximately 12 GB of memory, and need on the order of 10 minutes to calculate on a single processor. For $A_{\rm max}=125$, less than a GB of memory was needed and partition functions could be calculated in less than a minute. For hadron gases at temperatures of 150 MeV, $A_{\rm max}=250$ was sufficient for patch volumes $\lesssim 700$ fm$^3$. If multiple patch volumes are to be explored for the same temperature, computational time can also be saved by realizing that the partition functions scale as $\Omega^A$. Thus, if one performs a calculation for some initial volume $\Omega_0$, scaling can provide results for new volumes with minimal computation.

Once the partition function is calculated, event generation is remarkably fast. The time to generate an event scales linearly with the volume, or equivalently, linearly with the average number of particles generated. Running sufficient events to generate a million individual particles can be accomplished within a few seconds on a single CPU. Unlike Metropolis methods where events are modified by considering small changes to existing events, such as in \cite{Oliinychenko:2020cmr}, each event in this method is perfectly independent of previous events.
% !TEX root =  CCmoments.tex

\section{Bose and Fermi Statistics}\label{sec:bose}

Including Bose and Fermi statistics into the recursive relations for partition functions is straight-forward, and was shown in \cite{Cheng:2002jb,Pratt:1999ns}. The method is related to that used for calculating the effects of multi-boson interference for pion interferometry \cite{Pratt:1993uy}. In a fixed volume the partition function can be first treated as the usual procedure of accounting for $n$ identical particles being in different single-particle states. This includes the $1/n!$ term to account for the fact that the particles are indistinguishable, i.e. the Gibbs paradox. If $m_\ell$ indistinguishable particles are in the same single-particle state $\ell$, one must correct the weight by a factor of $m_\ell!$ for each level, which can also be thought of as the analog of the symmetrized relative wave function with all the momenta being equal. For fermions, the weight becomes $(-1)^{\ell-1}m_\ell$. As demonstrated in \cite{Pratt:1999ns}, the recurrence relation to the partition function then becomes
\begin{eqnarray}\label{eq:Zbf}
Z_{A}(Q,B,S)&=\frac{1}{A}\sum_h \sum_n Z_{A-n}(Q-nq_h,B-nb_h,Q-nq_h)z_{h,n}(\pm 1)^{n-1},
\end{eqnarray}
where $z_{h,n}$ is the partition function for $n$ particles in some level, 
\begin{eqnarray}
z_{h,n}&=\sum_\ell e^{-n\beta \epsilon_\ell},
\end{eqnarray}
and $\ell$ refers to single-particle levels of energy $\epsilon_\ell$. The $\pm 1$ refers to bosons or fermions. For hadron gases in the high temperature environments of relativistic heavy ion collisions, only pions have a non-negligible correction from quantum degeneracy. The correction of order $n$ for any level $\ell$ is of the order $e^{-\beta \epsilon_\ell}$ lower than the previous term. This factor is largest for zero momentum, and for pions becomes $e^{-\beta m}$, where $m$ is the pion mass. For the zero-momentum level at a temperature of 150 MeV, the factor is $e^{-m/T}\approx 0.4$, and as the system cools the factor falls slightly \cite{Greiner:1993jn}. For a more characteristic thermal momentum the factor is $\approx 0.1$. For heavier particles the factor is always small in the context of relativistic heavy ion collisions. For example, for a zero-temperature $\rho$ meson the factor is a fraction of a percent.

Given that symmetrization is only being applied to pions, which are bosons, one can incorporate these corrections into the Monte Carlo procedure outlined in Sec. \ref{sec:theoryMC}. For fermions this might be problematic because of the negative weights coming from the $(-1)^{n-1}$ factors in Eq (\ref{eq:Zbf}), but fortunately this is unneccessary because the degeneracy of fermions is negligible in the systems considered here. For pions, the algorithm is adjusted by treating each value of $n$ as being a different species, with charges $nq_h$ and with the partition function calculated with a reduced temperature $T\rightarrow T/n$. If one picks such a species in step 3 of the algorithm, $n$ pions are generated, all with the same momentum. For a finite system, the $n$ pions would be assigned small relative momenta on the order of the inverse system size.

It is well known that bosonic effects can broaden multiplicity distributions, consistent with negative binomial distributions \cite{Carruthers:1983my,Carruthers:1989jj}, making them super-Poissonian. One of the goals of this study is to discern how bosonic statistics alter the skewness and kurtosis. 
% !TEX root =  CCmoments.tex

\section{Models with Uniform Efficiency and Fixed Charge}\label{sec:uniformeff}

Even for a volume of fixed charge, finite efficiency and acceptance leads to non-zero fluctuations. The degree to which these fluctuations affect the skewness and kurtosis was worked out in \cite{Savchuk:2019xfg} for emission of a fixed charge where the probability of any charge being observed was a constant $\alpha$. If $\alpha$ were zero or unity, there would be no fluctuations, and because the charge on those particles that are not observed must fluctuate exactly opposite to the charge that is observed, the even moments must be symmetric about $\alpha=1/2$, and the odd moments must be anti-symmetric. One of the most important results of \cite{Savchuk:2019xfg} is that for fixed $Q$ and $\alpha$ the ratios of cumulants depend only on $\alpha$ and the variance and mean of the underlying multiplicity distribution. Even though $Q$ is fixed, the net number of charged particles $M$ can fluctuate. 

From \cite{Savchuk:2019xfg}, the probability that $M$ charged particles with total charge $Q$ will result in a measured charge $q=n_+ - n_-$ due to a uniform efficiency $\alpha$ is the convolution of two binomial distributions
\begin{eqnarray}
P(q|M,Q)&=&\sum_{n_+=0}^{(M+Q)/2}\sum_{n_-=0}^{(M-Q)/2}\frac{[(M+Q)/2]![(M-Q)/2]!}{[(M+Q)/2-n_+]![(M-Q)/2-n_-]!n_+!n_-!}
\alpha^{n_++n_-}(1-\alpha)^{M-n_+-n_-}.
\end{eqnarray}
After some tedious algebra, one can find the cumulants for fixed multiplicity,
\begin{eqnarray}
C'_1&=&\overline{q}=\alpha Q,\\
\nonumber
C'_2&=&\langle (q-\overline{q})^2\rangle_M = \alpha(1-\alpha)M,\\
\nonumber
C'_3&=&\langle (q-\overline{q})^3\rangle_M = \alpha(1-\alpha)(1-2\alpha)Q,\\
\nonumber
C'_4&=&\langle(q-\overline{q})^4\rangle_M -3\langle(q-\overline{q})^2\rangle_M^2\\
\nonumber
&=&\alpha(1-\alpha)-6\alpha^2(1-\alpha)^2M.
\end{eqnarray}
Here, the primes emphasize that the averages $\langle\cdots\rangle_M$. denote that they consider only those events with fixed base multiplicities $M$. The even moments are all linear in $M$, while the odd moments are linear in $Q$. It might appear that due to the linearity one might replace $M$ with $\langle M\rangle$ for a fluctuating base multiplicity, but the second term in the fourth cumulant
\begin{eqnarray}
-3\sum_M P(M) \langle(q-\overline{q})^2\rangle_M^2 &\ne&
-3\langle(q-\overline{q})^2\rangle^2.
\end{eqnarray}
Instead,
\begin{eqnarray}
\langle(q-\overline{q})^2\rangle^2&=&\alpha^2(1-\alpha)^2\overline{M}^2.
\end{eqnarray}
This provides a contribution to $C_4$, and the cumulants and their ratios become \cite{Savchuk:2019xfg}
\begin{eqnarray}
\label{eq:savchuk}
C_1&=&\alpha Q,\\
\nonumber
C_2&=&\alpha(1-\alpha)\overline{M},\\
\nonumber
C_3&=&\alpha(1-\alpha)(1-2\alpha)Q,\\
\nonumber
C_4&=&\alpha(1-\alpha)\overline{M}-6\alpha^2(1-\alpha)^2\overline{M}\\
\nonumber
&&+3 \alpha^2 (1-\alpha)^2\langle(M-\overline{M})^2\rangle,\\
\nonumber
\frac{C_2}{C_1}&=&(1-\alpha)\frac{\overline{M}}{Q},\\
\nonumber
\frac{C_3}{C_1}&=&(1-\alpha)(1-2\alpha),\\
\nonumber
\frac{C_3}{C_2}&=&(1-2\alpha)\frac{Q}{\overline{M}},\\
\nonumber
\frac{C_4}{C_2}&=&1+3\alpha(1-\alpha)(\omega_M-2),\\
\nonumber
\omega_M&\equiv&\frac{\langle(M-\overline{M})^2\rangle}{\overline{M}}.
\end{eqnarray}
The relative variance of the multiplicity of the base distribution, $\omega_M$, is unity for a Poissonian distribution. In that case $C_3/C_1$ and $C_4/C_2$ fall below unity for non-zero $\alpha$. The assumptions for this relation are that the charge and volume are fixed, that the efficiency is uniform, and that Bose and Fermi symmetrization is ignored. This relation is important as it implies that if one understands the efficiency and the second moment of the base multiplicity distribution, one can construct a baseline of cumulant ratios, and attribute any deviation from the baseline as due to fluctuations in $Q$, which is precisely the goal.

A reasonable value of $\alpha$ can be taken from balance function analysis. If a charge is observed, there should be a balancing charge emitted nearby, and detected with probability $\alpha$. This should correspond to the integrated strength of the charge balance function, which for electric charge is in the neighborhood of 0.35 assuming the full acceptance of the STAR TPC. Of course, $\alpha$ is not a constant. If a charge is observed in the center of the detector, its balancing charge has a better chance of being observed than for one observed near the periphery of the acceptance. Nonetheless, for the purposes of roughly setting expectations, this suggests that $C_4/C_2$ and $C_3/C_1$ can significantly differ from unity. In the next five subsections, \ref{sec:volumefluc} thrugh \ref{sec:hadrongas_cheap}, we consider how various effects, aside from fluctuating the base charge $Q$, might push $\omega_M$ to be either super-Poissonian, $\omega_M>1$ or sub-Possonian, $\omega_M<1$. We discuss the effects of volume fluctuations, charge conservation with a single type of charge, decays, Bose condensation, and finally the sensitivity to considering a realistic collection of resonances accounting for all three types of conserved charge.

\subsection{Volume Fluctuations}\label{sec:volumefluc}

Heavy ion experiments measure collisions spanning a range of impact parameters. Even for a fixed impact parameter, energy deposition might significantly vary depending on how many nucleons actually collided or how many jets were produced. If more energy is deposited into a fixed volume, it might expand further before it hadronizes, resulting in larger volumes when the system hadronizes. Experimental analyses attempt to minimize these fluctuations by constraining a given fluctuation measurement to a specific centrality bin, where ``centrality'' might be defined by multiplicity, transverse energy, or energy deposition in a forward calorimeter. To reduce auto-correlation, centrality measurements are usually constrained to particles other than those used to construct the moments. Nonetheless, it is inevitable that a range of initial conditions is explored within any centrality bin, and one might thus expect the base multiplicity distribution to broaden. If the patch volumes are fixed at some constant value, volume fluctuations can be thought of as a fluctuation in the number of patches. 

It is tempting to expect that the ratio of cumulants would be independent of the number of patches, and thus impervious to volume fluctuations. Independently, each cumulant $C_n$ scales linearly with the number of patches because there are no cross correlations between patches. Thus,
if $P_p(N)$ is the probability of having $N$ patches, and if $c_n$ is the cumulant for a single patch, the cumulant for the overall system is
\begin{eqnarray}
\label{eq:patchprob}
C_n&=&\sum_NP_p(N) Nc_n,
\end{eqnarray}
and the ratio of cumulants is 
\begin{eqnarray}
\frac{C_n}{C_m}&=&\frac{\sum_NP_p(N)Nc_n}{\sum_MP_p(M)Mc_m}\\
\nonumber
&=&\frac{c_n}{c_m}.
\end{eqnarray}
Thus, the ratio of cumulants is independent of the number of patches, and so independent of the overall volume, though the ratio can still depend on the volume of an individual patch.

Unfortunately, although the ratios of cumulants are independent of the overall volume, they are not impervious to fluctuations of the overall volume, or equivalently, to fluctuations of the number of patches \cite{Gazdzicki:1992ri,Gorenstein:2011vq,Gazdzicki:2013ana}. To illustrate this, one can consider a system with a distribution $P_p(N)$  as in Eq. (\ref{eq:patchprob}), where the average number of patches is $\overline{N}$. For $n>1$ the cumulants are expressed in terms of $\delta Q=Q-\overline{Q}$. Because $\overline{Q}$ varies with the number of patches we rewrite it as
\begin{eqnarray}
\delta Q&=&\delta Q_N+\delta N\overline{Q}_1,\\
\nonumber
\delta Q_N&\equiv&Q-N\overline{Q}_1,
\end{eqnarray}
where $\overline{Q}_1=c_1$ is the average charge in a single patch. Here $\delta Q_N$ is the fluctuation of the charge for a specific number $N$ of patches, and $\delta N=N-\overline{N}$. Inserting these definitions into the expressions for the cumulants,
\begin{eqnarray}\label{eq:volfluc}
C_1&=&\kappa_1c_1,\\
\nonumber
C_2&=&\sum_NP_p(N) \langle(\delta Q_N+\delta N\overline{Q}_1)^2\rangle=\kappa_1c_2+\kappa_2c_1^2,\\
\nonumber
C_3&=&\sum_NP_p(N) \langle(\delta Q_N+\delta N\overline{Q}_1)^3\rangle=\kappa_1c_3+\kappa_3c_1^3,\\
\nonumber
C_4&=&\sum_NP_p(N) \langle(\delta Q_N+\delta N\overline{Q}_1)^4\rangle-3C_2^2\\
\nonumber
&=&\kappa_1c_4+3\kappa_2c_2^2+6\kappa_3c_2c_1^2+\kappa_4c_1^4,
\end{eqnarray}
where $\kappa_n$ are the cumulants of $P_p(N)$,
\begin{eqnarray}\label{eq:kappadef2}
\kappa_1&=&\sum_N NP_p(N)=\overline{N},\\
\nonumber
\kappa_2&=&\sum_N (N-\overline{N})^2P_p(N),\\
\nonumber
\kappa_3&=&\sum_N (N-\overline{N})^3P_p(N),\\
\nonumber
\kappa_4&=&\sum_N (N-\overline{N})^4P_p(N)-3\kappa_2^2.
\end{eqnarray}
If the number of patches is fixed, i.e. the overall volume does not fluctuate, then $\kappa_{n>1}=0$ and each cumulant satisfies $C_n=\overline{N} c_n$. The ratio of cumulants then cancels the factor $\overline{N}$. 
Once the volume fluctuates, the ratios of cumulants of the charge distribution depend on the ratios of cumulants of $P_p(N)$. Methods have been constructed to reduce the dependence on volume fluctuations \cite{Begun:2014boa,Gazdzicki:2013ana,Gorenstein:2011vq,Sangaline:2015bma}, and have been applied to STAR data \cite{Luo:2017faz,Esumi:2020xdo}. These methods are built on the assumption that the observable used to identify the volume is correlated to the fluctuating charge only through the fact that they both scale with the volume. If the phase space used for the centrality measure is clearly distinct of the phase space over which charge is measured, this should be a good approximation. 

Equations (\ref{eq:volfluc}) and (\ref{eq:kappadef2}) are built on the assumption that the average charge scales linearly with the number of patches. This differs from the assumptions going into Eq. (\ref{eq:savchuk}), where it was assumed that charge was fixed within the overall volume. If the total charge in the volume is absolutely fixed despite fluctuations in the number of patches, then volume fluctuations affect the answer in that the relative variance of the multiplicity distribution, $\omega_M$, is increased by volume fluctuations as described in Eq. (\ref{eq:savchuk}).

\subsection{Chemically Equilibrated Canonical Distribution}\label{sec:singlecharge}

\begin{figure}
\centerline{\includegraphics[width=0.5\textwidth]{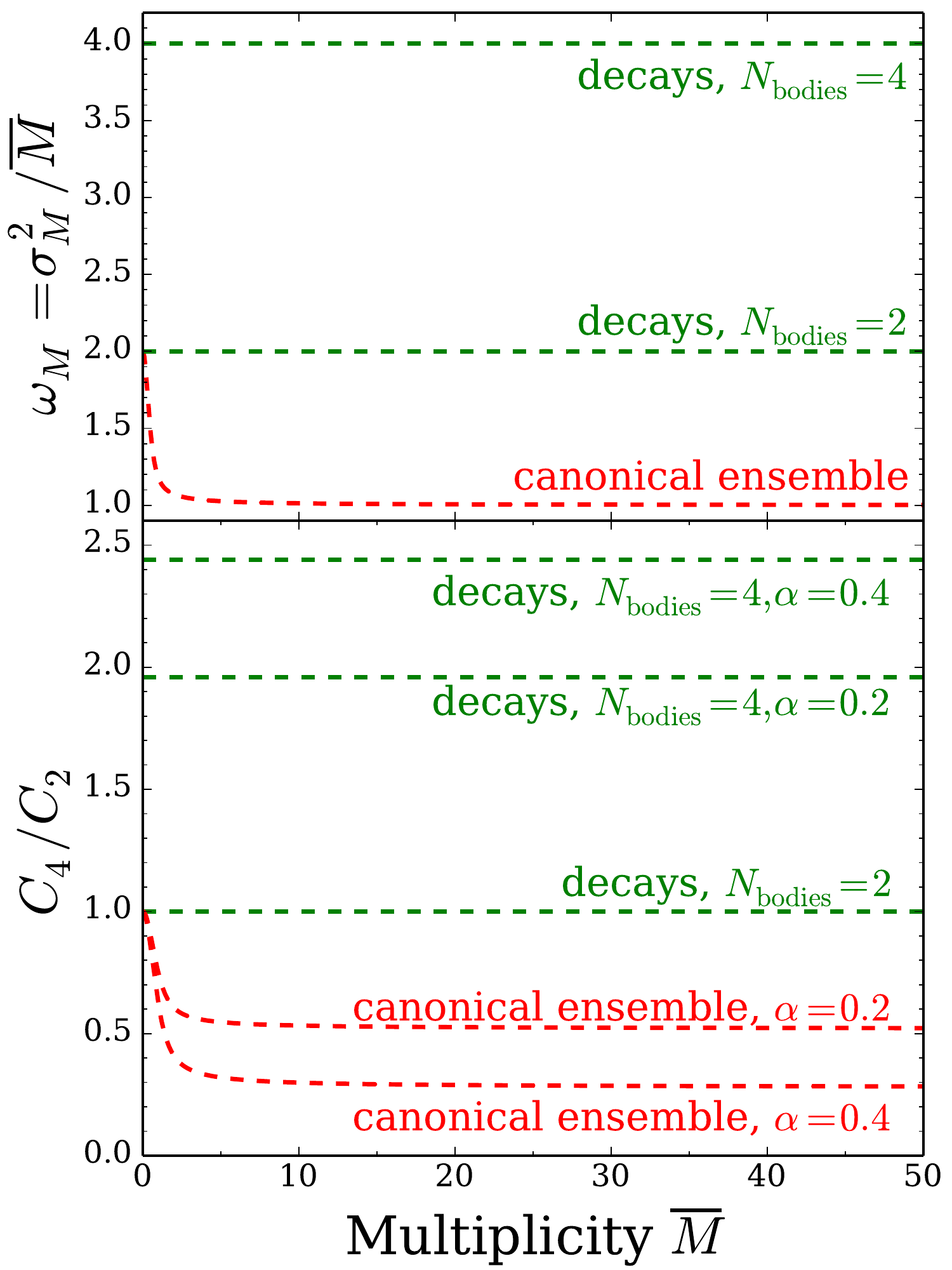}}
\caption{\label{fig:savchuk}
Upper panel: The relative variance, $\omega_M$, of the charged multiplicity distribution is shown for three cases for a system carrying only a single type of $\pm$unit charge. For a neutral equilibrated system (red dashed line) in a canonical ensemble, $\omega_M$ approaches unity, the Poissonian limit for higher mean multiplicities $\overline{M}$. In the low multiplicity limit the multiplicity will be either $M=0$ or $M=2$, which gives $\omega_M=2$. This is in contrast to a system where one has a Poissonian distribution of neutral particles which all decay into pairs (green dashed line) which gives $\omega_M=2$ for all multiplicities, or if the neutral particles all decay to four charges, which gives $\omega_M=4$.
Lower panel: According to Eq. (\ref{eq:savchuk}), which was derived in \cite{Savchuk:2019xfg}, the ratio $C_4/C_2$ is determined by $\omega_M$ and the acceptance probability $\alpha$. It is unity for $\omega_M=2$, below unity for $\omega_M<2$, and above unity for $\omega_M>2$. This shows that charge conservation in an equilibrated system pushes $C_4/C_2$ below unity, whereas if an equilibrated system of neutral particles decays, and if the decay products do not reform into the resonances, the resulting ratio of $C_4/C_2$ is unity for two-body decays, or above unity for four-body decays.}
\end{figure}

One can calculate the relative variance, $\omega_M$, for an equilibrated system of a single type of charge, where the measurement directly samples the equilibrated canonical distribution. This ignores decays, which in later stages proceed without the regeneration needed to maintain equilibrium. The results of the equilibrated canonical ensemble in Eq. (\ref{eq:PAgivenQcanonical}) can be used to find $\omega_M$, the relative variance of the multiplicity distribution, and thus generate the moments using Eq. (\ref{eq:savchuk}). For large average multiplicities $\overline{M}$, the value of $\omega_M$ approaches unity, the Poissonian limit. For small $\overline{M}$ it approaches two. This is expected, because for a fixed charge, you can only add particles pair-wise. In fact, as will be shown in the next section, if one generates a Poissonian number of pairs, the relative variance will be $\omega_M=2$ for all multiplicities. Because $\omega_M$  varies from from two to unity in a canonical ensemble, as the system size increases from zero to infinity, the values of $C_4/C_2$ and $C_3/C_1$ stay below unity as shown in Fig. \ref{fig:savchuk}.

The fact that $\omega_M$ is above unity for small systems (due to only being able to sample even numbers of charges) is known as canonical suppression. This lowers the mean multiplicity and raises the relative variance, $\omega_M$. As seen in Fig. \ref{fig:savchuk} it also raises the ratio $C_4/C_2$ relative to its value for a large system. 

\subsection{Decays}\label{sec:decays}

A system of uncorrelated neutral particles that decay to charged particles also leads to fixed zero net charge. But unlike the results for the equilibrated canonical ensemble above, this can lead to ratios of $C_4/C_2>1$. If each neutral particle decays into $N_{\rm bodies}$ charged particles, where the charges are $\pm 1$, the charged particle multiplicity distributions become 
\begin{eqnarray}
\langle M_{\rm ch}\rangle&=&N_{\rm bodies}\overline{M}_0,\\
\nonumber
\langle(M_{\rm ch}-\overline{M}_{\rm ch})^2\rangle&=&N_{\rm bodies}^2\langle(M_0-\overline{M}_0)^2\rangle,\\
\nonumber
\omega_M&=&\frac{\langle(M_{\rm ch}-\overline{M}_{\rm ch})^2\rangle}{\overline{M}_{\rm ch}}\\
\nonumber
&=&N_{\rm bodies}\frac{\langle(M_0-\overline{M}_0)^2\rangle}{\overline{M}_0}.
\end{eqnarray}
If the emission of neutrals is Poissonian, the result is simple, $\omega_M=N_{\rm bodies}$. Fig. \ref{fig:savchuk} illustrates how this picture affects $C_4/C_2$. In this case, because $\omega$ depends only on $N_{\rm bodies}$ and does not change with multiplicity or system size, the cumulants are also independent of multiplicity. For $N_{\rm bodies}=2$ the cumulant ratios do not even depend on the efficiency. For $N_{\rm bodies}>2$ the ratio $C_4/C_2$ exceeds unity, so if charge creation proceeded through the creation and decay of neutral clusters it would be easy to generate large values of $C_4/C_2$. This has been discussed at length in \cite{Bzdak:2018uhv}.

In an equilibrated system, decays and recombination have equal rates. However, as the system decouples recombination stops and decays proceed until only stable hadrons remain. During the hadronic phase the number of charged particles nearly doubles due to these decays. Thus, if the systems do equilibrate, then decay after chemical freeze-out, one would expect the ratio $C_4/C_2$ to lie somewhere between the value for the canonical ensemble in Fig. \ref{fig:savchuk} and the value for pure decays with $N_{\rm bodies}=2$. STAR's experimental results for net proton fluctuations indeed satisfy this expectation, but their measured $C_4/C_2$ for net charge fluctuations exceed unity, which can only be attained if $N_{\rm bodies}>2$. Very few hadronic decays proceed via more than one charged pair, but one could have decays of clusters or hot spots. Other possible explanations for having $C_4/C_2>1$ are volume fluctuations, which were discussed previously, or Bose symmetrization, which will be discussed ahead.

\subsection{Bose correlations}\label{sec:bose_uniform}

It has long been understood that Bose correlations induce super-Poissonian fluctuations \cite{Carruthers:1983my,Carruthers:1989jj}. Here, we illustrate how Bose effects combine with charge conservation to determine $C_4/C_2$. Partition functions for a gas of positive and negative pions, $m_\pi=139.57$ MeV/$c^2$, were considered to be kinetically equilibrated but with a chemical potential enforcing a fixed average density. If a gas is created at chemical equilibrium one expects $\mu=0$, but if it cools while maintaining a fixed number of pions, and if the pion number is fed by decays, an effective chemical potential, $\mu$, is required. This changes the average number of pions by adjusting the phase space density as
\begin{eqnarray}
f(\vec{p})&=&\frac{e^{-(E_p-\mu)/T}}{1-e^{-(E_p-\mu)/T}}.
\end{eqnarray}
Because the system is out of equilibrium, and because the number of positives and negatives are nearly the same, the effective chemical potential has the same sign for both $\pi^+$ and $\pi^-$. At decoupling one expects kinetic temperatures to fall near 100 MeV and the pion chemical potential to grow to perhaps as high as 75 MeV \cite{Greiner:1993jn}. This estimate can be understood by the fact that the phase space occupancies should stay roughly constant for fixed entropy per particle for an expansion at fixed entropy, which suggests a chemical potential of approximately 50 MeV by the time the system cools to 100 MeV. Decay products further feed the phase space occupancy. Pion condensation occurs when this chemical potential reaches the pion mass. In the absence of Bose effects, this requires roughly doubling the phase space density of pions as compared to the expectations above. If such conditions were realized Bose effects could result in super-radiance \cite{Pratt:1993uy}, which should be accompanied by large multiplicity fluctuations. 

\begin{figure}
\centerline{\includegraphics[width=0.4\textwidth]{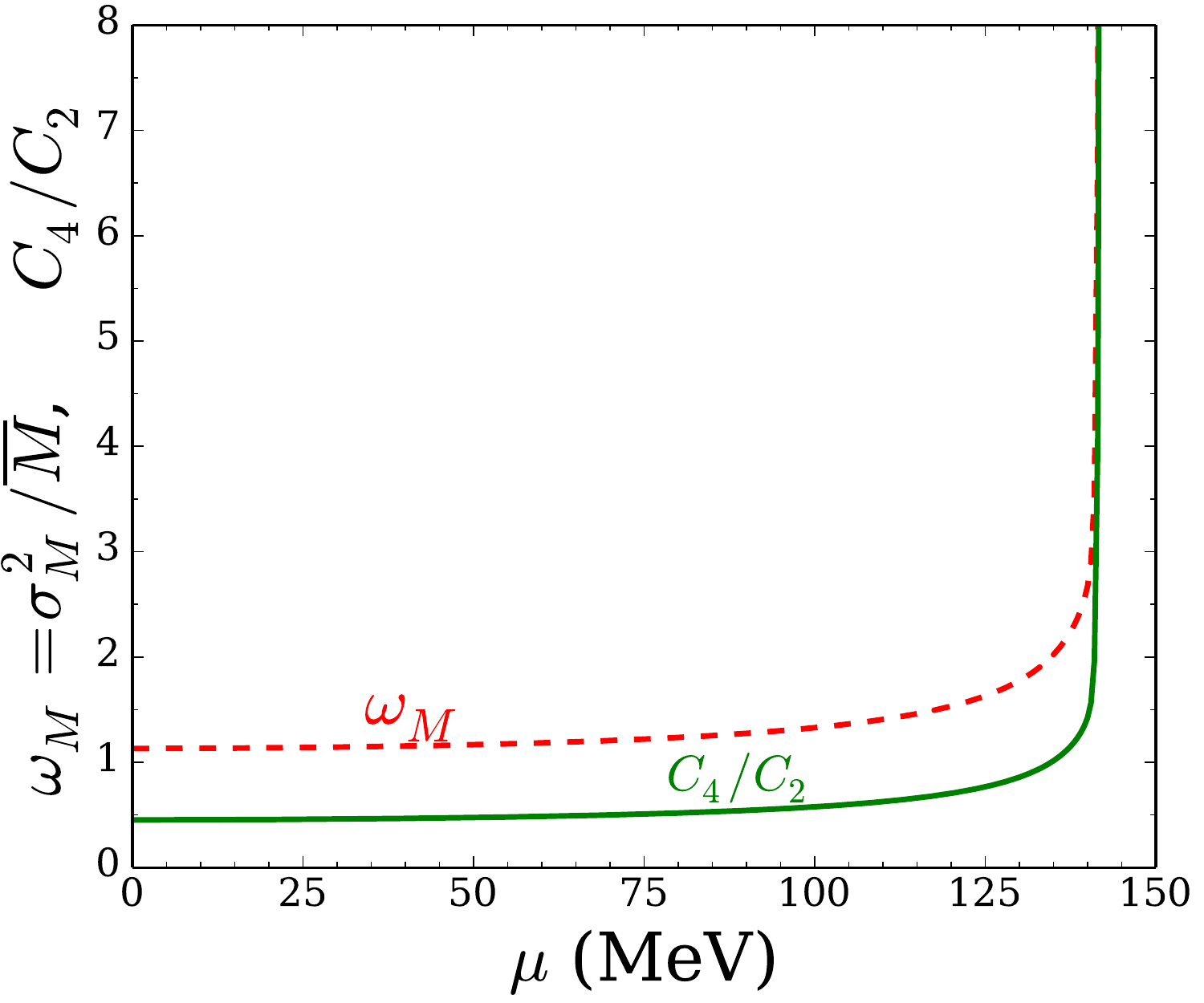}}
\caption{\label{fig:cheapbose}
Fluctuations for a canonical ensemble of pions at fixed charge, $Q=0$, with an effective chemical potential, $\mu$, applied to adjust the net pion number in a volume of 500 fm$^3$ at a temperature of 100 MeV. The relative variance of the multiplicity distribution (dashed red line) and the ratio $C_4/C_2$ of the net-charge distribution (solid green line) grow dramatically as $\mu$ approaches the pion mass. Super-radiant effects can take place once $\mu$ reaches $m_\pi=139.57$ MeV. Heavy-ion collisions are expected to decouple with effective chemical potentials near 75 MeV, which is well below the onset of large fluctuations.
}
\end{figure}
As a function of the effective chemical potential $\mu$, the partition function for a pion gas in a volume of 500 fm$^3$ and at a temperature of 100 MeV was calculated from Eq. (\ref{eq:Zbf}). For $\mu$ in the range of of 75 MeV the relative variance of the charged multiplicity distribution is $\omega_M\approx 1.2$, which modestly increases $C_4/C_2$ as shown in Fig. \ref{fig:cheapbose} for a fixed efficiency of $\alpha=0.3$. More dramatic results for $C_4/C_2$ are not expected unless the chemical potential reaches within a few MeV of the pion mass. As discussed above, this is not expected. However, if the number of pion sources fluctuated wildly from one event to another, and if there were some events with twice the number of sources emitting into the same phase space, super-radiance might occur in some small fraction of the events. Such behavior would strongly contradict expectations based on chemical equilibrium.

\subsection{Hadron Gas}\label{sec:hadrongas_cheap}

Thus far, all the simple examples presented in this section considered a system with one type of conserved charge, but in actuality a hadron gas obeys the conservation of three types of charge: baryon number, strangeness and electric charge. This invalidates the use of the recurrence relation of Eq. (\ref{eq:savchuk}) and requires the application of Eq. (\ref{eq:recurrence}), or if Bose corrections are included, Eq. (\ref{eq:Zbf}). Here, we calculate the canonical ensemble, $Z_A(Q,B,S)$, using the recurrence relation, then apply the Monte Carlo techniques of Sec. \ref{sec:theoryMC} to generate sample sets of particles. The calculation is based on a large number, $\sim 300$, of hadron resonances listed in \cite{Tanabashi:2018oca}. After generating the particles, unstable resonances are decayed. For this section, a uniform efficiency $\alpha$ is assumed, independent of species, momenta, or whether the products came from a weak decay (charged pions and kaons were not decayed).

Emission is assumed to come from sub-volumes, or patches, of fixed size. Because each sub-volume is independent, there are no correlations between the various sub-volumes. Also, because cumulants scale linearly with the number of sub-volumes, ratios of cumulants can depend only on the patch volume, not on the overall volume. However, if the number of patches fluctuates, the result is modified by volume fluctuations as shown in Sec. \ref{sec:volumefluc}.

Here, the sensitivity to three parameters is investigated. The three varied parameters are the patch volume $\Omega$, the baryon density $\rho_B$, and the fixed efficiency $\alpha$. When one parameter is varied, the other two are set at default values. The defaults are set at $T=150$ MeV, $\rho_B=0$, $\alpha=0.3$ and the default patch volume is $\Omega=200$ fm$^3$. The electric charge density is set to half of the given baryon density. After these sensitivities are studied, calculations at finite baryon density are modified so that the net baryon and electric charges responsible for the non-zero charge density are allowed to fluctuate across sub-volumes according to a Poissonian distribution. This is motivated by the fact that charges transported far away from mid-rapidity at early times cause this non-zero baryon density, so local charge conservation should play little role in the distribution of net charge across patches.

\begin{figure}
\centerline{\includegraphics[width=0.4\textwidth]{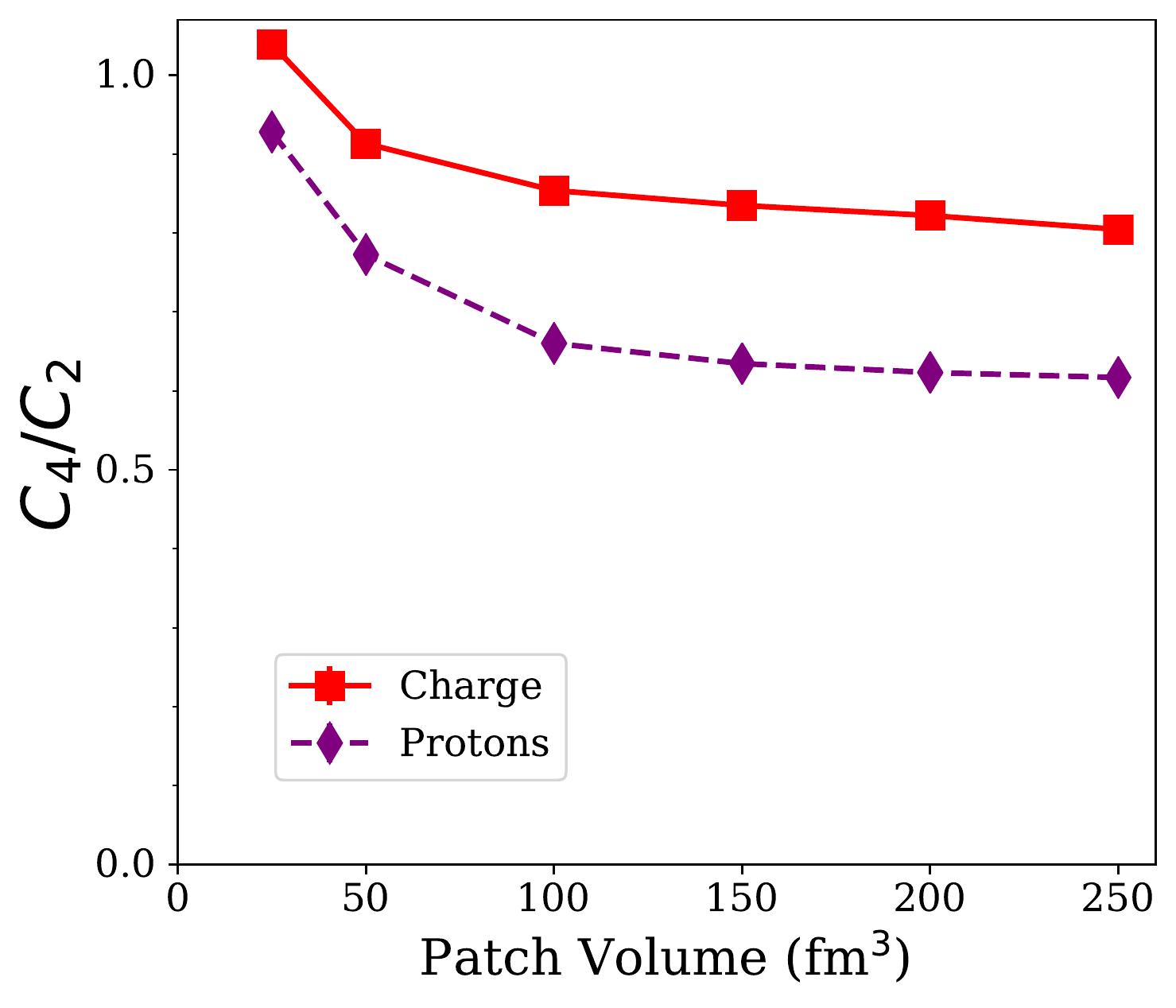}}
\caption{\label{fig:uniform_vs_volume}
The ratio of cumulants is shown for a $\rho_B=0$ system as a function of the patch volume. For small patches a baryon is always accompanied by an extra anti-baryon, but for larger systems the observation of a baryon might also enhance the probability that one fewer baryons is present in the remainder.}
\end{figure}

The dependence on patch size is exhibited in Fig. \ref{fig:uniform_vs_volume}. For small patches the emission of baryons is discouraged because each baryon must be accompanied by the existence of an anti-baryon. Thus, the Boltzmann factor, $e^{-M/T}$, for the mass is compounded by the necessity of having a second accompanying massive particle. Once the volumes become larger, and the mean number of baryon pairs exceeds unity, the observation of a baryon might be accompanied by seeing one fewer baryon in the remainder of the system, rather than seeing an extra anti-baryon. In fact, this is exactly what happens in a canonical ensemble in the large volume limit. In that limit, if a baryon is observed in some small amount of phase space, the mean number of baryons observed in the remainder of the system is decreased by one half, while the mean number of anti-baryons is increased by one half. It is expected that the dependence of cumulant ratios on patch size should disappear as the patch size increases beyond the threshold for the mean baryon density to be unity. Figure \ref{fig:uniform_vs_volume} displays this behavior by displaying the ratio $C_4/C_2$ as a function of patch volume. Because the mean multiplicity for charged particles exceeds unity at a smaller volume, the ratio levels off faster for net charge than for net protons. Qualitatively, the same behavior was seen for a system with a single type of charge in Fig. \ref{fig:savchuk}.

\begin{figure}
\centerline{\includegraphics[width=0.4\textwidth]{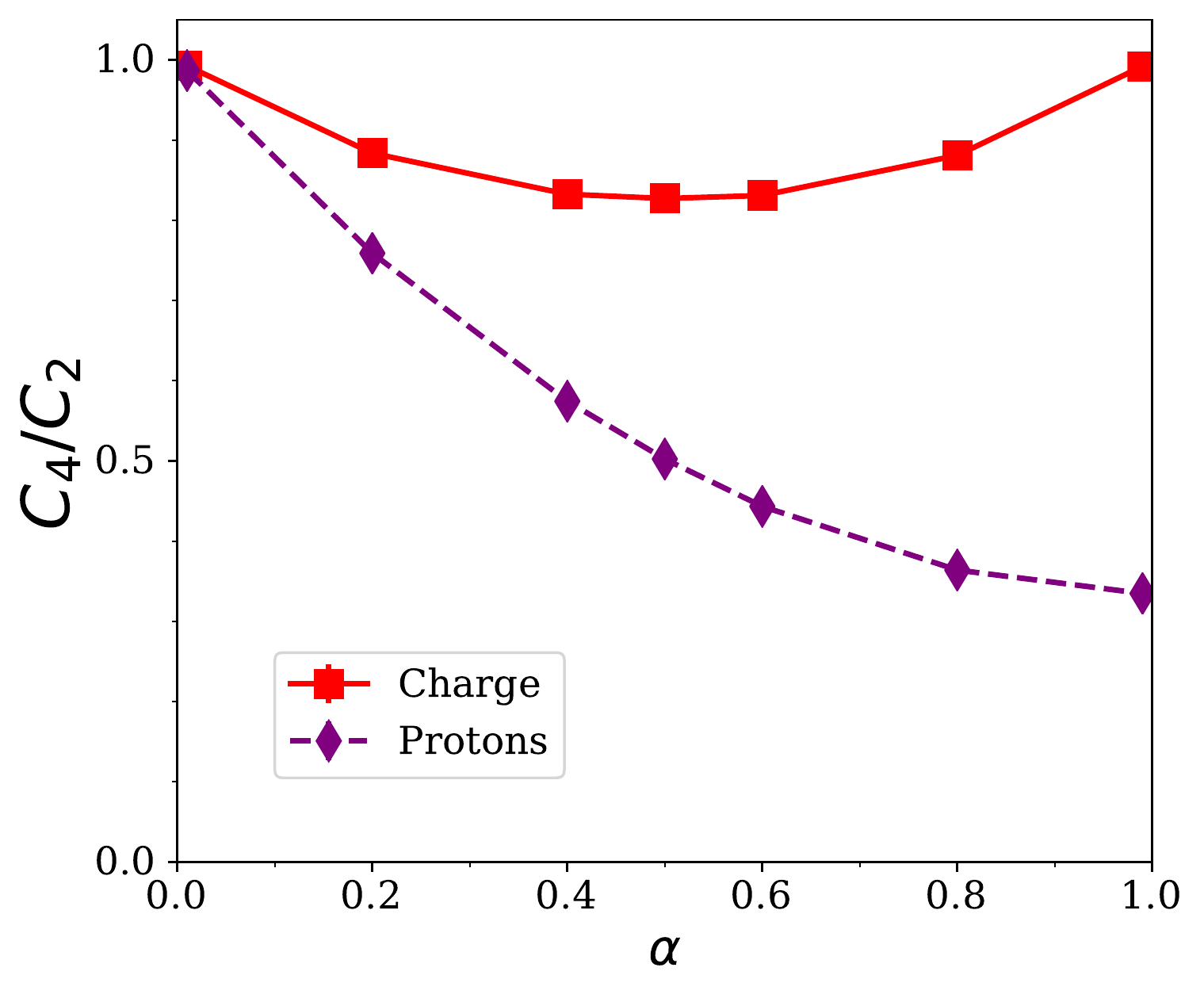}}
\caption{\label{fig:uniform_vs_alpha}
As the fixed acceptance probability $\alpha$ approaches zero, distributions become Poissonian, and the ratios $C_4/C_2$ approach unity. Even for perfect acceptance, the fluctuations are non-zero for a multi-charge system because the conserved charges carried by a proton can be balanced by an array of other species, and charge within the proton and anti-proton sector is not conserved. Thus, the ratio is not symmetric about $\alpha=0.5$ as one might infer from Eq. (\ref{eq:savchuk}), because protons and anti-protons represent only a fraction of the species that carry electric charge and baryon number.}
\end{figure}

The sensitivity to the acceptance probability, $\alpha$, is displayed in Fig. \ref{fig:uniform_vs_alpha}. As $\alpha$ approaches zero, distributions tend to become Poissonian, because the number of particles measured is dominantly either zero or unity. Eq. (\ref{eq:savchuk}) describes similar behavior for a system of a single type of charge. For net charge, the ratio is symmetric about $\alpha=0.5$. This is because when net charge is conserved, any charges not measured fluctuate exactly opposite to those that are measured. The same property is apparent in Eq. (\ref{eq:savchuk}). However, the distribution of net protons considers only protons and anti-protons, so it does not represent the entirety of particles that carry baryon number or electric charge. Thus, that distribution is not symmetric about $\alpha=0.5$ and fluctuations are large even for $\alpha \rightarrow 1$.

\begin{figure}
\centerline{\includegraphics[width=0.4\textwidth]{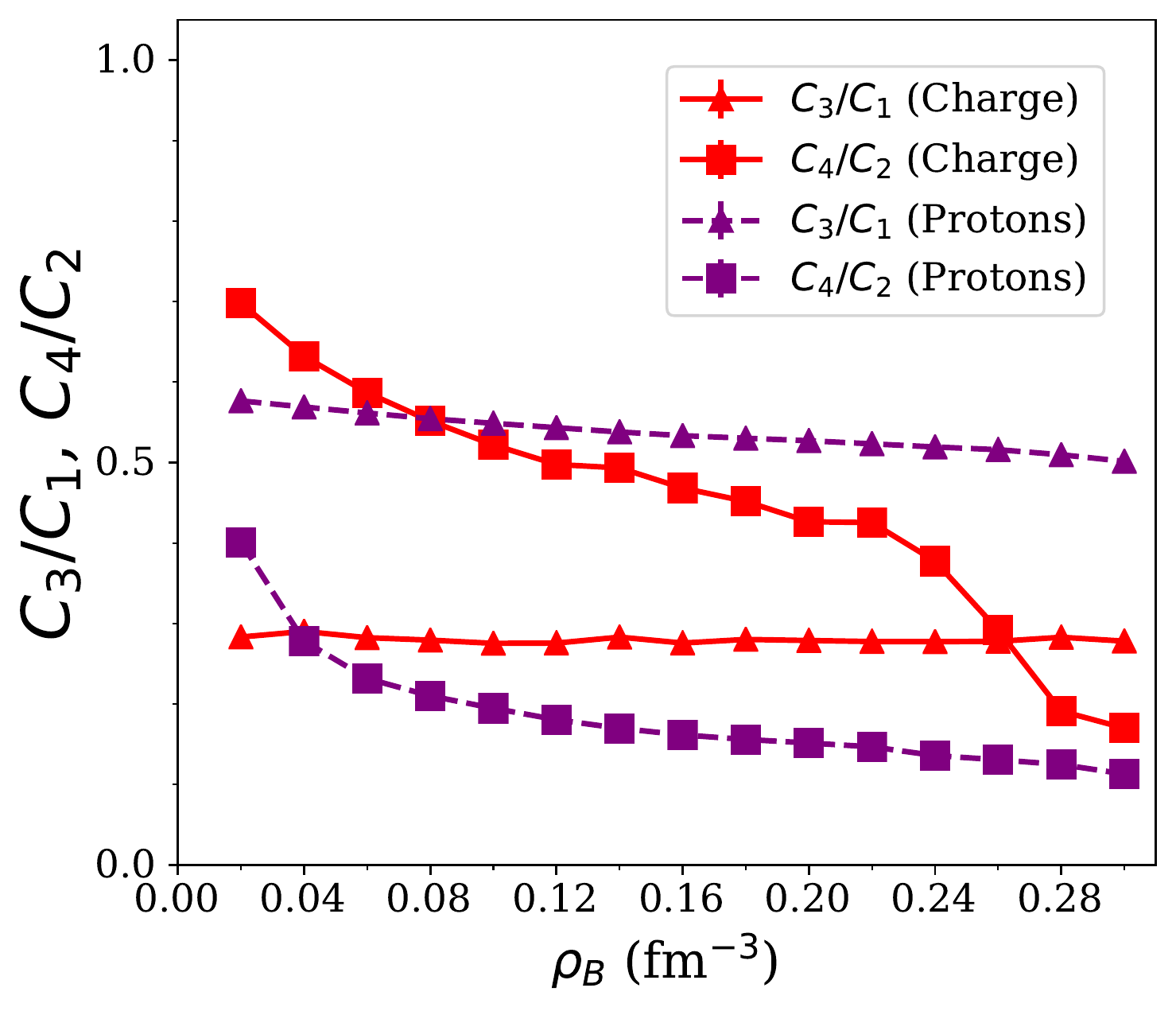}}
\caption{\label{fig:uniform_vs_rho_fixedQ}
The ratios $C_4/C_2$ and $C_3/C_1$ are shown for net charge and for net baryon number as a function of the baryon density. The ratios fall increasingly below the Poissonian limit of unity as the baryon number is increased. For this calculation the net baryon charge was fixed at $B=\rho_BV$ and the net electric charge was fixed at $Q=B/2$.}
\end{figure}

Figure \ref{fig:uniform_vs_rho_fixedQ} presents $C_4/C_2$ and $C_3/C_1$ as a function of baryon density. The patch volume is fixed at 200 fm$^3$ and the densities chosen correspond to a fixed number, $4,8,12, \cdots$, of baryons. The choice is made to display $C_3/C_1$ rather than $C_3/C_2$ because $C_3/C_1$ would be unity for uncorrelated emission. The ratios all decrease, moving further from unity, as baryon density increases. The densities displayed cover the range of what might be reached in heavy-ion systems at the point when temperatures fall below 150 MeV.

\begin{figure}
\centerline{\includegraphics[width=0.4\textwidth]{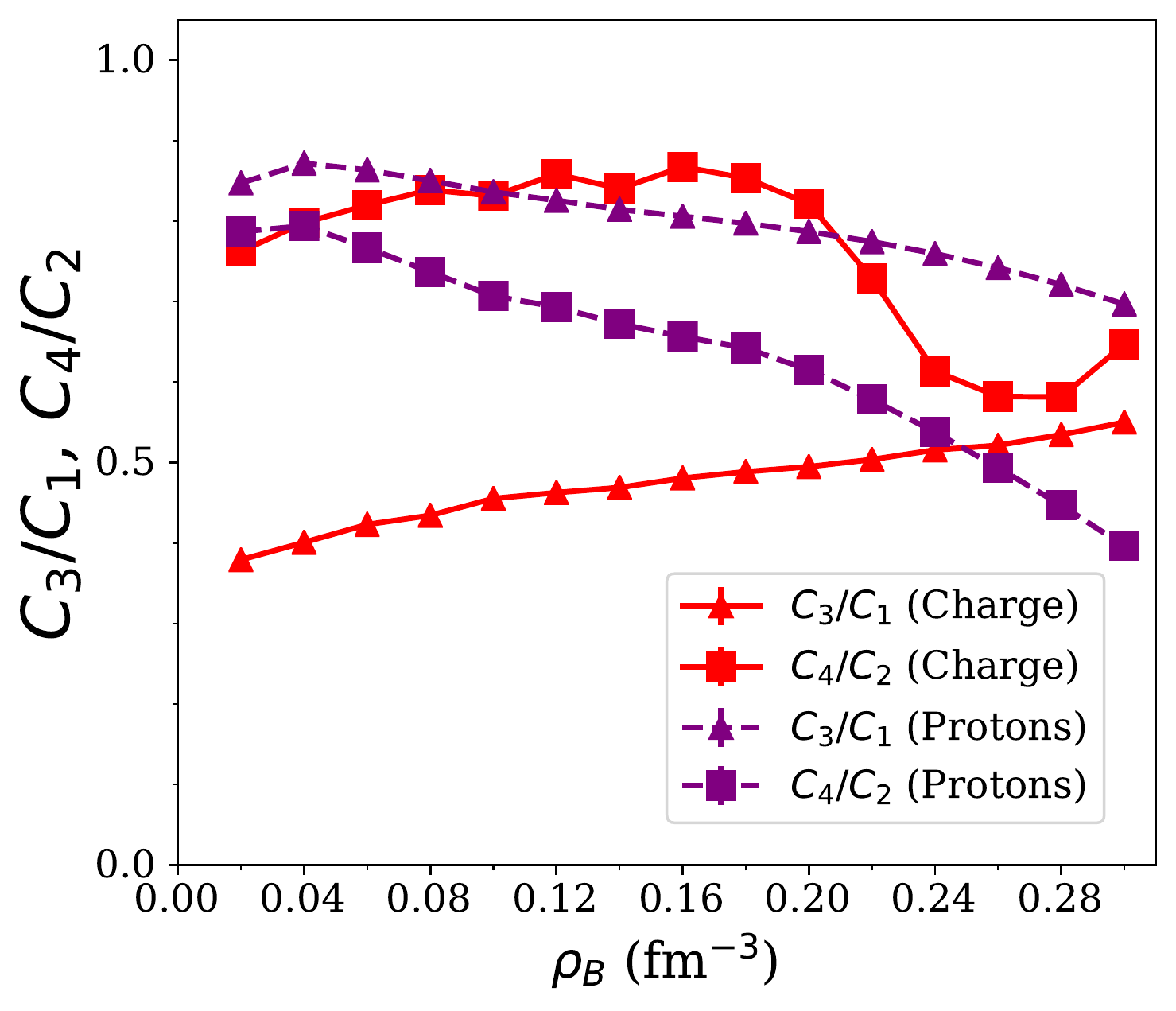}}
\caption{\label{fig:uniform_vs_rho_fluctuatingQ}
The ratios $C_4/C_2$ and $C_3/C_1$ are shown for net charge and for net baryon number as a function of the baryon density. Calculations differ from those in Fig. \ref{fig:uniform_vs_rho_fixedQ} in that the net charge fluctuates according to a Poissonian distribution. The ratios rise relative to those in the fixed-charge case, but they remain below unity.}
\end{figure}

For the calculations illustrated in Fig. \ref{fig:uniform_vs_rho_fixedQ}, the net baryon number, electric charge, and strangeness were all fixed. Because of local charge conservation and the limits of diffusion, balancing charges created after the collision are constrained to stay within the same neighborhood. This constraint is adjusted by setting the size of the patch volume. However, even for measurements at mid-rapidity, non-zero charges can arise due to the transfer of baryon number and electric charge from the projectile and target rapidities. These intruder charges are not typically balanced by charges within the acceptance of the detector, and thus their fluctuation should be considered separately. Figure \ref{fig:uniform_vs_rho_fluctuatingQ} shows how $C_4/C_2$ and $C_3/C_1$ behave as a function of baryon density, if the non-zero baryon number and electric charge are chosen to fluctuate within the patch according to a Poissonian distribution. As expected, the ratios rise as compared to the fixed case shown in Fig. \ref{fig:uniform_vs_rho_fixedQ}, but they do not exceed unity. Discerning these fluctuations of the actual net charges is of particular interest. These results suggest that the starting point for the ratios $C_4/C_2$ and $C_3/C_1$ is below unity for random uncorrelated net charges accompanied by an ensemble of baryon charges.

% RACHEL -- make figures for
% Default T=150 MeV, Patch Volume=200 fm^3, alpha=0.3
% A. C_4/C_2 for net protons and for net charge with default+ rho_B=0 vs Volume
% B. C_4/C_2 for net protons and for net charge with default+ rho_B=0 vs alpha
% C. C_4/C_2 and C_3/C_1 for net protons and for net charge for default vs rho_B=0={0,0.02,0.04,0.06..., 0.3} with fixed charge
% D. Same as (C) but with "fixed" charge fluctuating according to Poisson about average
%
% !TEX root =  CCmoments.tex

\section{Blast Wave Model with a Full Hadron Gas and Comparison to STAR Results}\label{sec:blast}

The calculations of the previous section were based on a simple picture, where each sub-volume emitted particles whose probability of being observed was uniform, denoted by $\alpha$. In practice, this probability depends on where the sub-volume is located within the overall reaction volume. A sub-volume in the region with spatial rapidity, $|\eta_s|>1$, emits particles that are only rarely recorded by detectors that specialize in mid-rapidity measurements. Even for a sub-volume centered at mid-rapidity, thermal motion can spread some of its emission to rapidities outside the acceptance. This is especially true when the experiments narrow their acceptance. For example, STAR's acceptance nominally covers $\pm 1$ units of pseudo-rapidity, but the acceptance for identified particles is confined to $\pm 0.9$ units. For real rapidity, the effective acceptance is narrowed further, due to the fact that the true rapidity $y$ is less than the pseudo-rapidity $\eta$. This difference is magnified for more massive or slower particles. In fact, STAR analyses of identified particles often enforce cuts that only consider particles with true rapidities $-0.5<y<0.5$. Of course, even particles within the rapidity acceptance must exceed some minimum transverse momentum, and the efficiency for being detected is imperfect. For particles identified only by charge, the efficiencies are typically $\gtrsim 80\%$, and for identified particles the efficiency falls by another few tens of percent.

Collective flow plays a critical role. First, longitudinal flow is what allows the measurement of rapidity to correlate to a measurement in coordinate-space rapidity. In the Bjorken model of boost-invariant flow, the mapping is simple in that a fluid element with spatial rapidity $\eta_s$ moves with a rapidity $\eta_s$, and particles emitted from a sub-volume with rapidity $\eta_s$ have rapidities $y\approx\eta_s$. This simple mapping is smeared by thermal motion. Collective radial flow and cooling combine to better align the spatial rapidity $\eta_s$ and the measured rapidity $y$. The thermal spread for pions is $\approx 0.6$ units of rapidity, while that for protons is $\approx 0.25$.

The extent of the region over which particles are emitted in spatial rapidity, $\eta_s$, should affect the result. If the extent in $\eta_s$ is small, there is an enhanced probability that a charge and its balancing charge will both be identified and cancel one another when assigning the net charge for an event. If the region extends over a large rapidity range, the effects of charge balance are minimized because there is a better chance that one charge will be observed while the other is outside the acceptance. The extent of the region in the transverse direction is less important. For emitting regions at the edge of the fireball, which have more collective velocity, there is a modest increase of having balancing charges both pushed into the acceptance. In addition to the size of the region over which particles from a given sub-volume are emitted, the overall size of the sub-volume at the point where chemical freeze-out occurs also plays a role as it sets the degree to which canonical suppression affects the results. As discussed in the previous section, this matters only for for smaller sub-volumes. 

For the model in this section, canonical sub-volumes are overlaid onto a blast-wave parametrization of collective flow. A filter is applied to the calculations, representing the experimental STAR acceptance and efficiency. This should be sufficiently realistic to make meaningful comparisons to STAR data. The four main parameters for the blast-wave model describe the radial flow $u_\perp$, the kinetic freeze-out temperature, $T_k$, and the baryon chemical potential and temperature at chemical freeze-out, $\mu_c$ and $T_c$. The chemical freeze-out temperature $\mu_c$ and chemical potential, $T_c$, are chosen to fit relative particle yields, while $T_k$ and $u_\perp$ are determined by simultaneously fitting the spectra of species with varying mass, typically $\pi,K,p$. A variety of parameterizations exist, such as having the velocity increase linearly from the origin, or the transverse rapidity, or having a sharp cutoff in radius vs. assuming a Gaussian profile. Depending on the choice, the value of $T_k$ and $u_\perp$ vary, but are typically in the neighborhood of $T_k\approx 100$ MeV and $u_\perp\approx 0.6$. For increasing multiplicities, the reaction volumes can expand and cool further, which leads to modestly increased values of $u_\perp$ and modestly decreased values of $T_k$ for either more central or for more energetic collisions. For this study the chemical freeze-out parameters were taken from \cite{Das:2012yq,Kumar:2012np}, which extracted $T_c$ and $\mu_c$ for a variety of beam energies. For this section, Bose statistics were ignored. Over 300 hadron species were included in the analysis. The collective flow velocities were chosen to reproduce the mean transverse momenta of both pions and protons.

The distribution of spatial rapidities $\eta_s$ over which all particles are emitted is chosen to be Gaussian, with a width $\sigma_0$ that depends on beam energy as
\begin{eqnarray}
\frac{dN}{d\eta_s}&\sim& e^{-\eta_s^2/2\sigma_0^2},\\
\nonumber
\sigma_0&=&0.4~y_{\rm beam},
\end{eqnarray}
where $\pm y_{\rm beam}$ are the rapidities of the incoming beams. This choice reproduces the rapidity widths measured at RHIC to roughly 10\% accuracy \cite{Flores:2016mtp}. The distribution of spatial rapidities for particles of a given sub-volume was also spread according to a Gaussian with a width $\sigma_\eta$. The parameter $\sigma_\eta$ describes how far particles created from the same sub-volume may have separated by the time of emission. For increasingly larger values of $\sigma_\eta$, the chance that any two observed particles are correlated decreases. The rapidity of a sub-volume was then distributed according to a Gaussian with width $\sqrt{\sigma_0^2-\sigma_\eta^2}$, so that the emission overall was characterized by the width $\sigma_0$. Whereas the other blast-wave parameters were chosen to describe spectra and yields, the parameter $\sigma_\eta$ is related to charge conservation. If charge is created early, and especially if the diffusion constant is large, the width might be close to one unit of spatial rapidity, whereas if all the charge were to come after hadronization, the width might more likely be a few tenths. Blast-wave models of charge balance functions, which are also driven by charge conservation and diffusion, suggest widths should be of the order of a third, but variations of a factor of two were not ruled out \cite{Schlichting:2010qia}. One goal of this section is to investigate the sensitivity to $\sigma_\eta$.

For this blast-wave model, each sub-volume was also assigned a radial transverse velocity according to a Gaussian,
\begin{eqnarray}
\frac{dN}{d^2u_\perp}&\approx&e^{-(u_x^2+u_y^2)/2u_\perp^2},
\end{eqnarray}
and a small spread in $\vec{u}_\perp$. However, given that for this study, cuts are not being considered in transverse momenta or azimuthal angle, varying $\vec{u}_\perp$ has little effect. Particles from each differential volume element were assigned momenta using the Monte Carlo algorithm described in \cite{Pratt:2010jt}.

The calculations of this section are filtered through a simplified model of the STAR detector's acceptance and efficiency. For unidentified particles, pseudo-rapidities are required to be between $\pm 0.5$ and transverse momenta are constrained to being above 200 MeV/$c$ and below 2 GeV/$c$. For identified particles, rapidities were restricted to being between $\pm 0.5$, and transverse moment were required to be between 200 MeV/$c$ and 1.6 GeV/$c$ for pions or kaons, and between 400 MeV/$c$ and 2 GeV/$c$ for protons and antiprotons. Because the STAR data are corrected for efficiency, the efficiencies were set to unity. In order to compare results to measurement from STAR, the baryon densities and chemical freeze-out temperatures were mapped to beam energy according to the analysis of \cite{Kumar:2012np}, which extracted chemical freeze-out temperatures and chemical potentials by considering ratios of particle yields. The parameter $u_\perp$, which controls the transverse radial flow, was chosen along with the kinetic breakup temperature, $T_k$, to simultaneously roughly fit the mean transverse momenta of both the protons and pions reported \cite{Abelev:2009bw}. The kinetic freeze-out parameters were chosen according to the parameterization,
\begin{eqnarray}
T_k&=&120-20f~{\rm MeV},\\
u_\perp&=&0.5+(0.74-0.5)f,\\
f&=&\frac{1}{\ln(2)}
\ln\left[ 1+ \frac{\sqrt{s}-\sqrt{s}_0}{\sqrt{s}_f-\sqrt{s}_0}\right],
\end{eqnarray}
where $\sqrt{s}_0=7.7$ GeV and $\sqrt{s}_f=200$ GeV. Decays were simulated, and the products of weak decays (aside from pions or charged kaons) were included in the analysis. Undoubtedly, a more realistic model of the acceptance might change the ratios, but given that these are ratios, and that the overall efficiency and acceptances are not wildly off, a more rigorous model of the acceptance is unlikely to change the result by more than a few percent.

The summation over sub-volumes was performed by summing over 400 values of $u_\perp$ and $\eta_s$ consistent with the distributions described above. For each value, emission of $N_{\rm sample}=2\times 10^5$ sub-volumes was simulated so that cumulants could be calculated for each value of $\eta_s$ and $u_\perp$. Because emission from different sub-volumes is uncorrelated, the cumulants for the total emission are simply the sum of the cumulants of each sub-volume. Further, because the net cumulants behave linearly in $N_{\rm sample}$, the ratios of net cumulants is independent of $N_{\rm sample}$.

\begin{figure}[htb]
\centerline{
\includegraphics[width=0.32\textwidth]{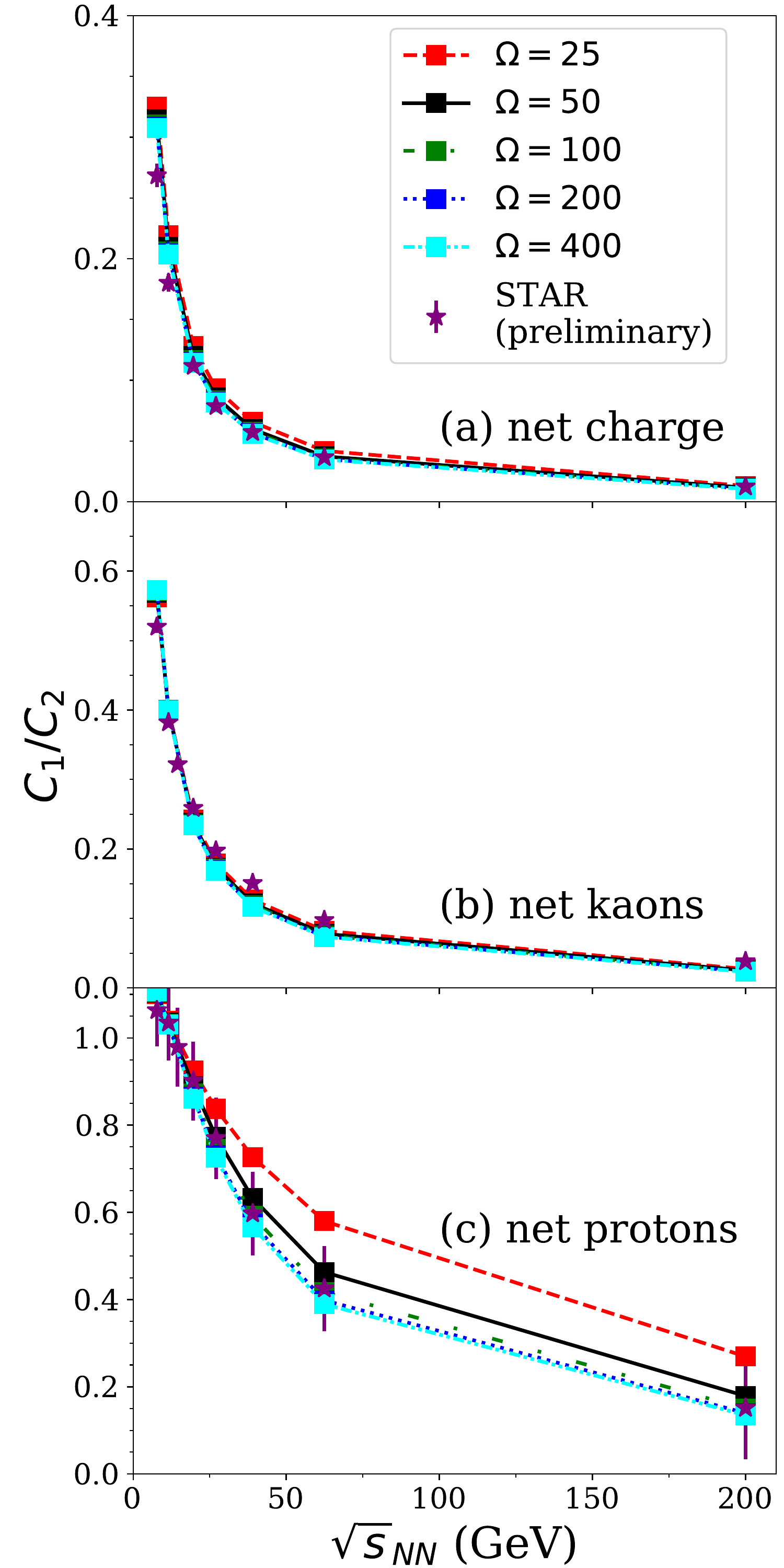}\hspace{0.02\textwidth}
\includegraphics[width=0.32\textwidth]{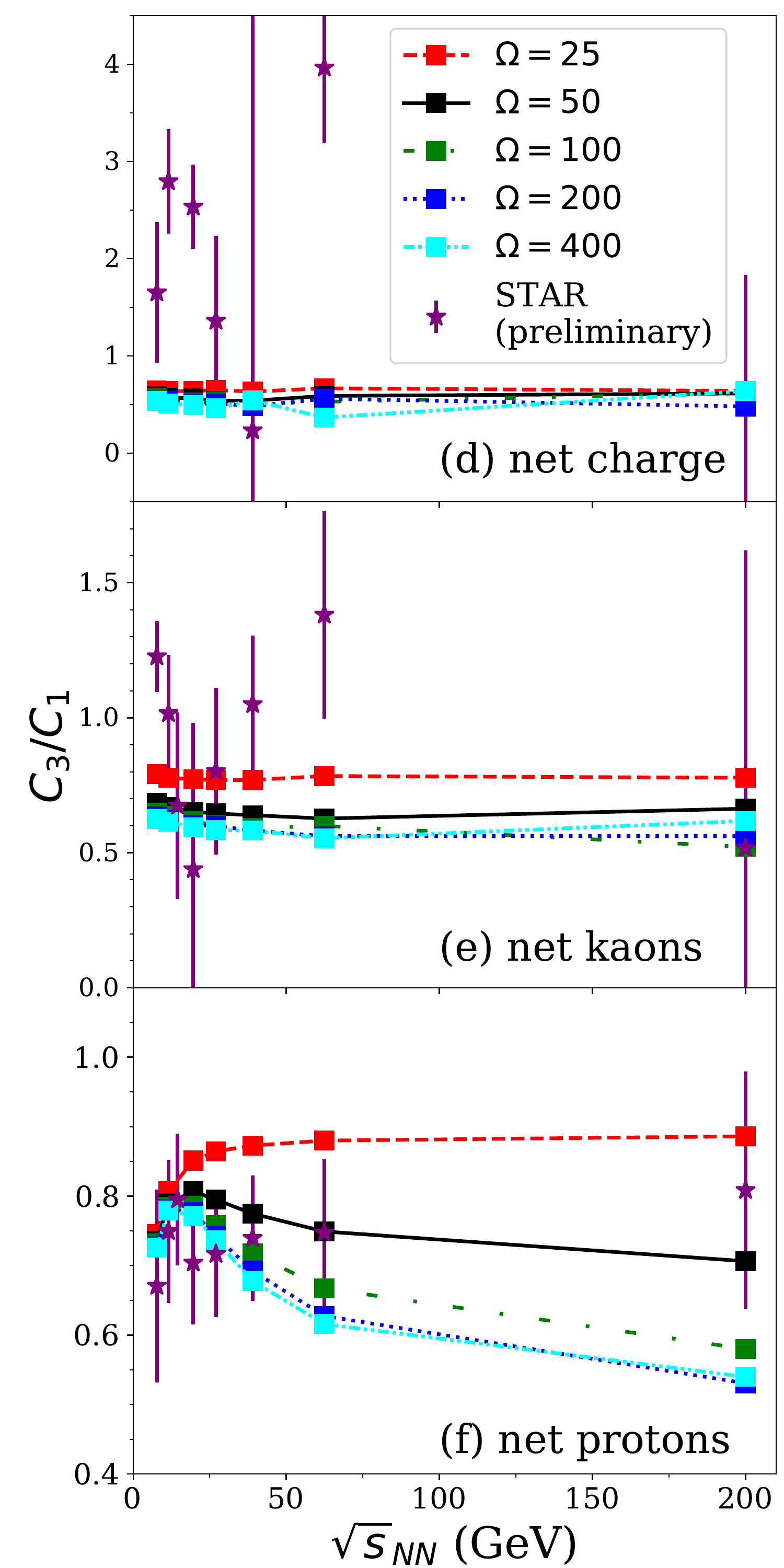}\hspace{0.02\textwidth}
\includegraphics[width=0.32\textwidth]{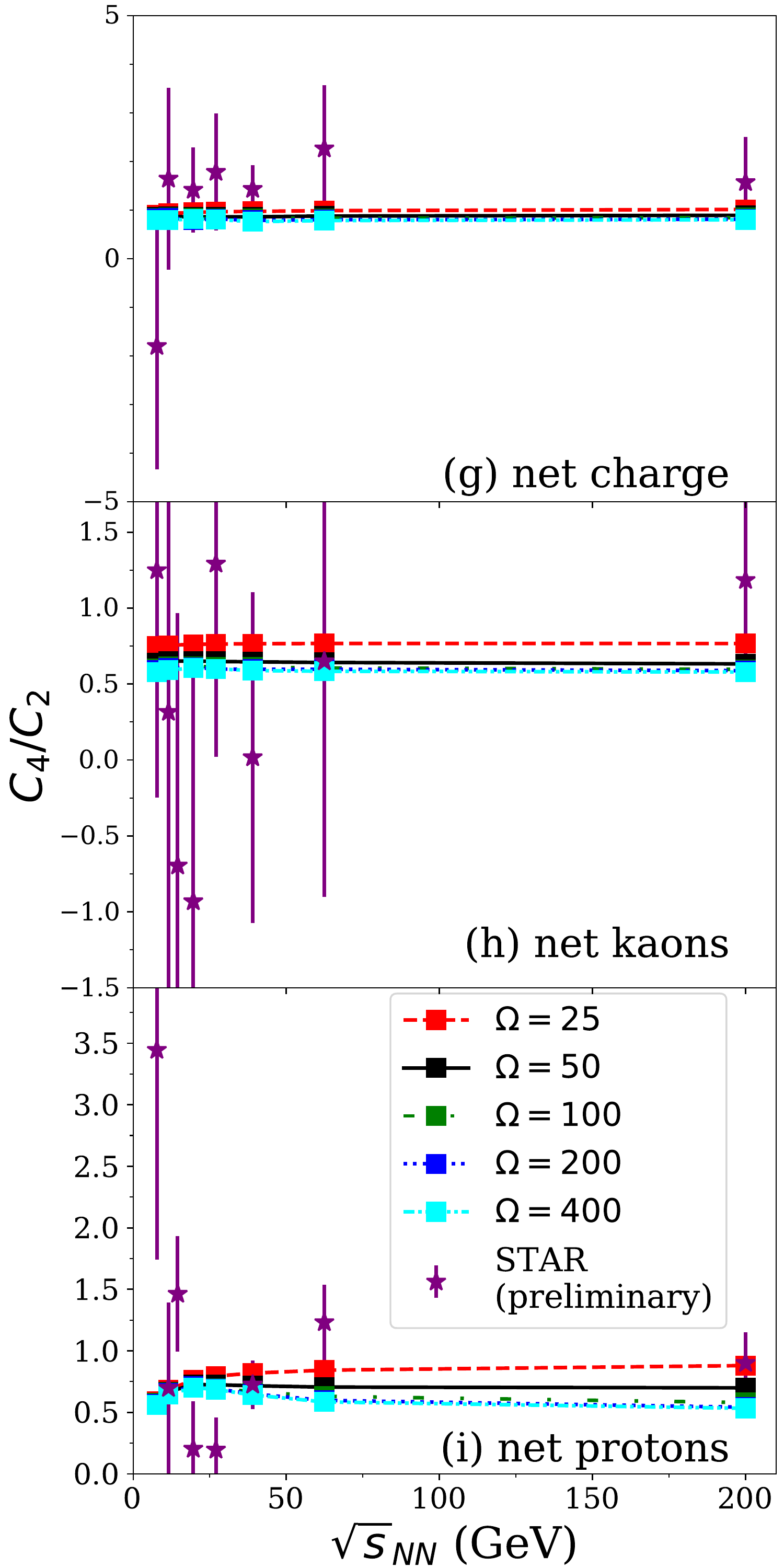}
}
\caption{\label{fig:bw_vs_omega}(color online) Ratios of cumulants from blast wave models, begining with $C_1/C_2$ (panels {\it a} - {\it} c), then skewness (panels {\it d} - {\it f}) and finally, kurtosis (panels {\it g} - {\it i}), are plotted for for different values of the sub-volume $\Omega$. For smaller sub-volumes the ratios $C_1/C_2$, $C_3/C_1$ and $C_4/C_2$ increase. For net charge, the ratios approach an asymptotic value once $\Omega$ begins to pass $\sim 50$ fm$^3$, whereas for net protons the ratios appear to approach the limit at somewhat higher values of $\Omega$. STAR measurements for net protons are not wholly dissimilar to the blast-wave calculations here, but those for net charge differ greatly. Additional physics from volume fluctuations might explain how $C_4/C_2$ might exceed unity, but it is difficult to explain how this might happen for net-charge distributions while leaving $C_4/C_2$ of the net-proton distribution unchanged.}
\end{figure}
Figure \ref{fig:bw_vs_omega} illustrates how results are sensitive to the size of the sub-volume, $\Omega$. This is the effective volume at which charge conservation is enforced at chemical freeze-out. For these calculations $\sigma_\eta$ was fixed at 0.3. For a volume of 100 fm$^3$ the average number of hadrons is several dozen. For small sub-volumes, where in a grand canonical ensemble the typical number of charges would be zero or one, the thermodynamic cost of having a second charge to balance the first charge reduces the probability of having any charges. As mentioned earlier, this is known as canonical suppression, and it lowers both the multiplicities and the moments. For larger sub-volumes, this thermodynamic cost vanishes as the system might have had fewer balancing charges of the same sign, rather than an extra balancing charge of the opposite sign. The characteristic volume for the disappearance of canonical suppression is the volume where the mean number of pairs exceeds unity. In general, for charged particles this sets in at around 50 fm$^3$, but for baryons the characteristic volume is closer to 100 fm$^3$ because of their being heavier and fewer. Thus, the proton moments for volumes of 50 fm$^3$ and 200 fm$^3$ differ more noticeably. The time for a fluid element to expand and cool to the point where it reaches chemical freeze-out tends to be on the order of 5 fm/$c$. These times are shorter for matter on the periphery, at a lower beam energy or centrality. The times are longer for fluid elements at the center, at higher beam energy, or in a more central collision. The maximum transverse distance a charge can travel before chemical freeze-out is ~$\approx 10$ fm if they move in opposite directions, and if charge moves diffusively or if the charge is created later in the reaction, the separation should be significantly less. As discussed below, charge can separate further along the beam axis, and because that is not well understood, the size of the sub-volume carries a large uncertainty. Anywhere from 50 fm$^3$ to a few hundred fm$^3$ might be reasonable. 

\begin{figure}
\centerline{
\includegraphics[width=0.32\textwidth]{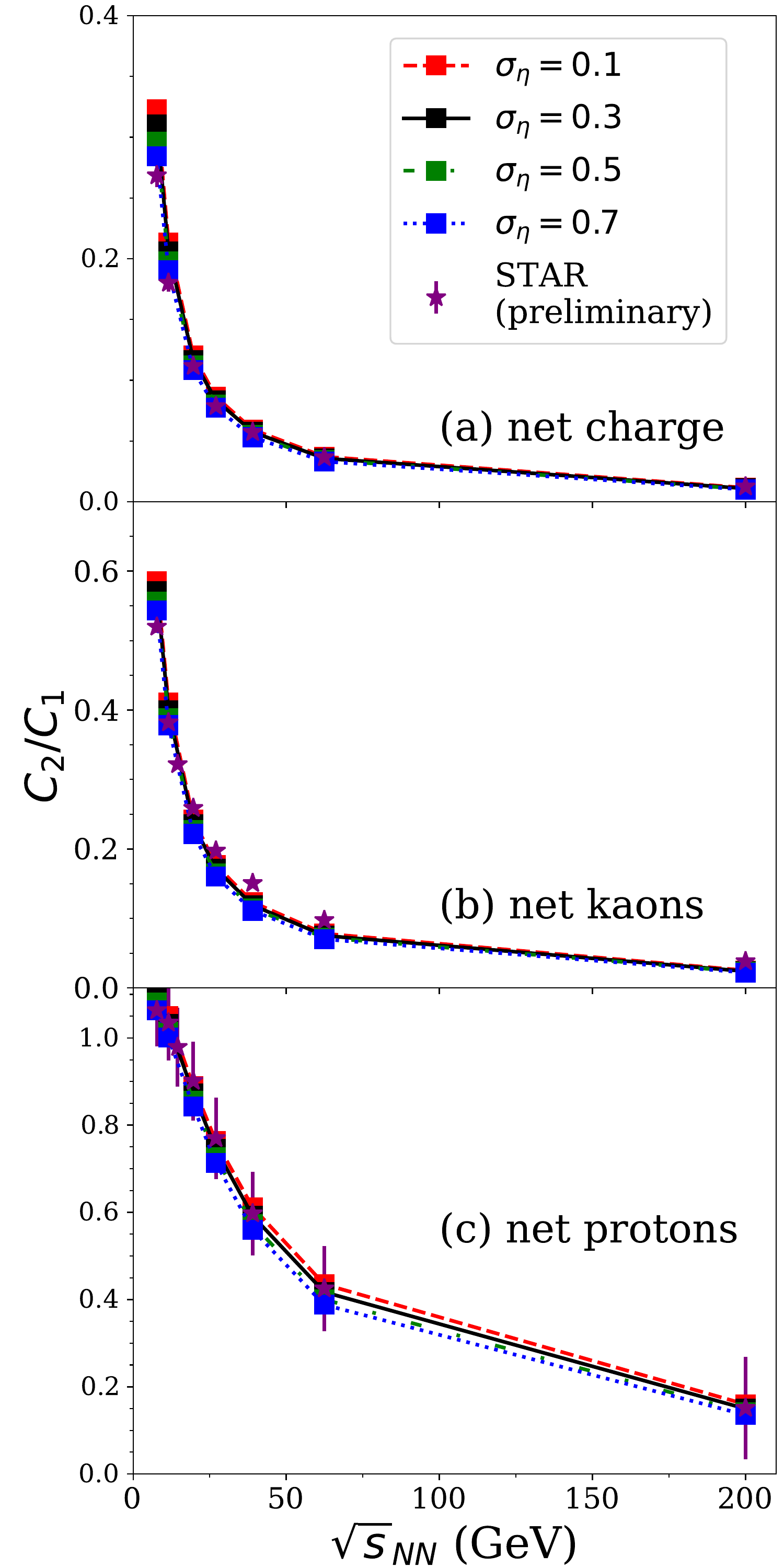}\hspace{0.02\textwidth}
\includegraphics[width=0.32\textwidth]{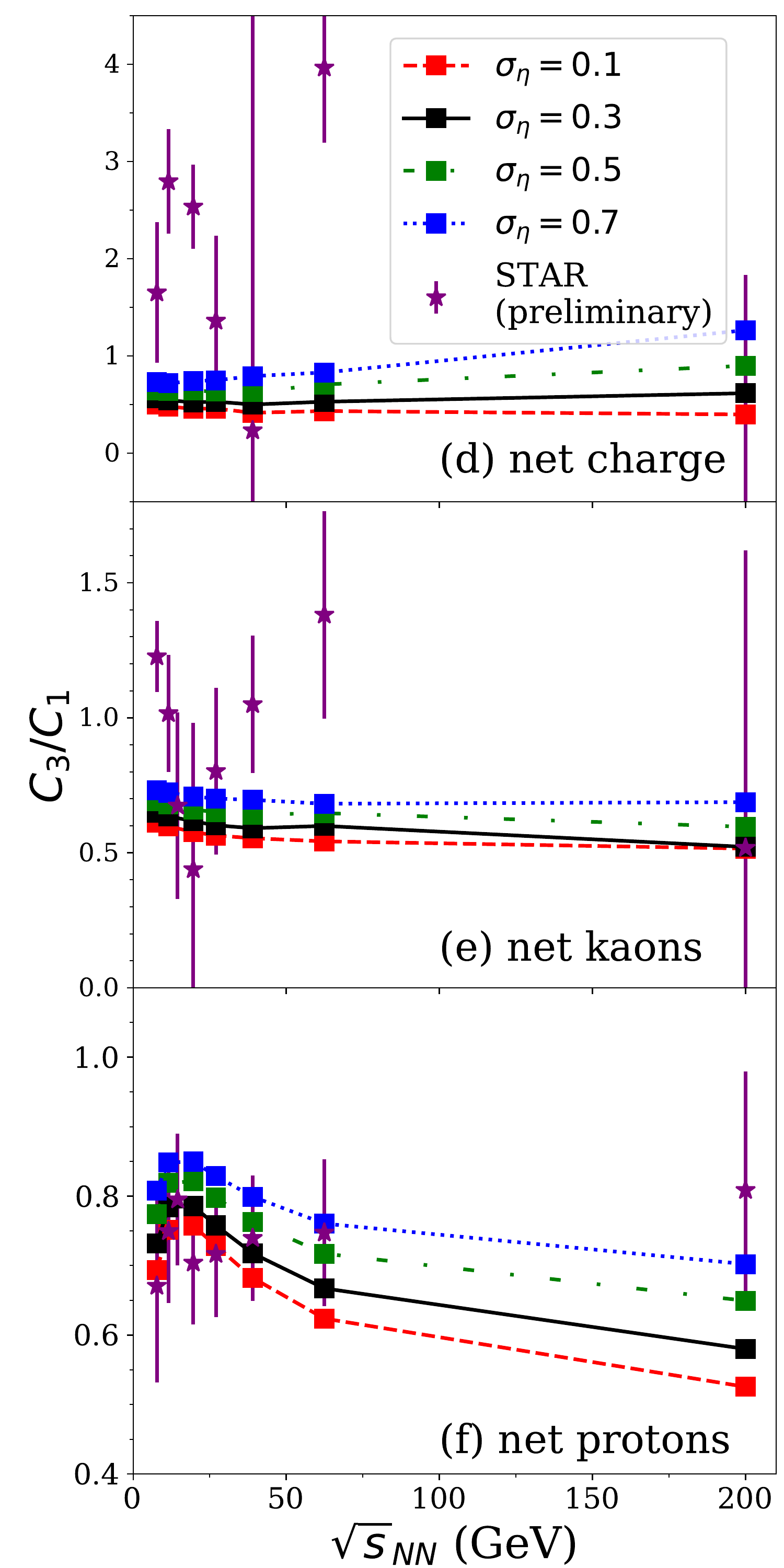}\hspace{0.02\textwidth}
\includegraphics[width=0.32\textwidth]{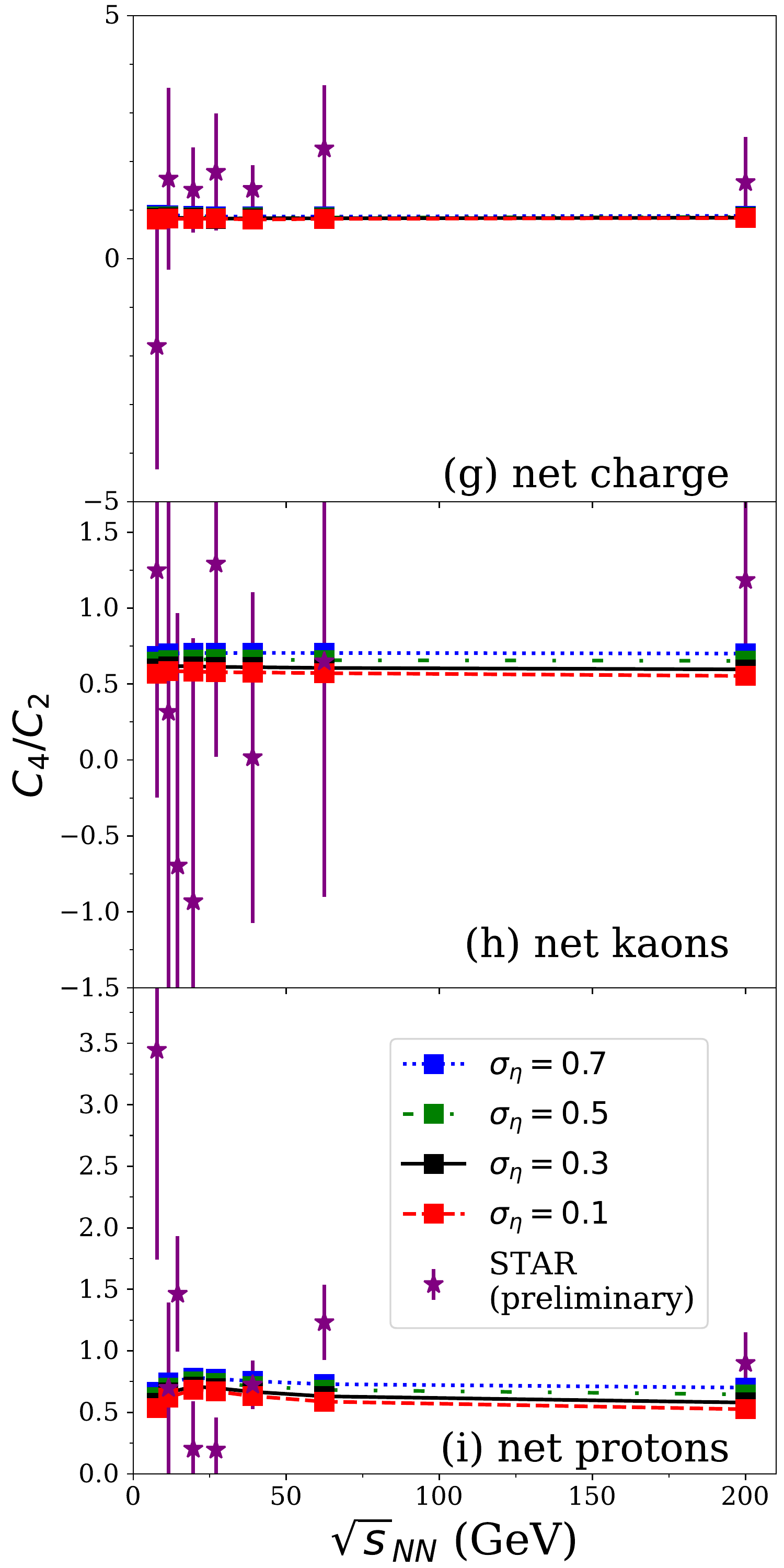}
}
\caption{\label{fig:bw_vs_sigmaeta} (color online) Ratios of moments are displayed for different values $\sigma_\eta$, which sets the longitudinal size of over which sub-volumes emit charge. A modest sensitivity is found for net protons, while the moments for net charge were fairly insensitive.}
\end{figure}

Balancing charges separate from one another, and depending on when they are created, they can diffuse apart from one another. This separation, represented by the parameter $\sigma_\eta$, accounts for the separation of balancing charges both before and after hadronization, or both before and after chemical freezeout. After chemical freeze-out charges might separate and mix between the sub-volumes. Because some of the separation comes after chemical freeze-out, this distance might exceed the scales representing the size of the sub-volume $\Omega$. Charge can spread further in the longitudinal direction due to the strong initial longitudinal collective flow at early times. This enhancement to the separation depends sensitively on when charge pairs are created. Matter thermalizes at an early time, where large collective velocity gradients along the beam axis are expected. If the motion is diffusive, the separation along the beam axis depends logarithmically on the ratio of the final time to the initial creation time \cite{Bass:2000az}.  Thus, if a pair is created at 0.2 fm/$c$, the separation for times, $0.2<\tau<1.0$ fm/$c$ is as important as the additional separation they gain during the times $1.0<\tau<5.0$ fm/$c$. For this reason the size of charge spread, $\sigma_\eta$, in spatial rapidity might be anywhere between a few tenths of a unit of rapidity to a full unit. Figure \ref{fig:bw_vs_sigmaeta} shows the sensitivity of the moments to this parameter. For large $\sigma_\eta$ the observation of a charge is less likely to influence the observation of a second charge, which is similar to having a lower efficiency. For low efficiencies one expects the behavior to be more Poissonian, and $C_4/C_2$ and $C_3/C_1$ to be closer to unity. Indeed, this is the case, but the dependence on $\sigma_\eta$, as shown in Fig. \ref{fig:bw_vs_sigmaeta}, is negligible for net charge and modest for net protons.

The beam energy dependence mainly derives from the fact that the baryon density is higher for lower beam energies. Given that $C_1$ falls with increasing beam energy, while $C_2$ increases because multiplicity increases, it is no surprise that the ratio $C_1/C_2$ falls with increasing beam energy. The skewness, $C_3/C_1$, should approach unity if emission is random, i.e. a Skellam distribution. The effects of local charge conservation keep $C_3/C_1<1$. In the limit that there are no anti-baryons, which is the limit of high net baryon density and equivalently low beam energy, the assumption that baryons are deposited amongst the sub-volumes according to a Poissonian distribution also drives $C_3/C_1$ for net protons closer to unity. Combined with higher canonical suppression for heavier particles, the sensitivity to $\Omega$, shown in Fig. \ref{fig:bw_vs_omega}, or to $\sigma_\eta$, illustrated in Fig. \ref{fig:bw_vs_sigmaeta}, is more pronounced for blast-wave calculations of net protons than of net kaons or of net charge.

The measure of kurtosis, $C_4/C_2$, varies only modestly with beam energy. For net charge, the ratio does not vary far from 0.8 in model calculations. In contrast, $C_4/C_2$ from model calculations for net protons varies from 0.6 to 0.9 depending on the values of $\Omega$ and $\sigma_\eta$. The models also exhibit a modest dependence on beam energy for $C_4/C_2$. The ratio rises approximately 20 percent as the beam energy increases to 20 GeV, then plateaus and falls 10\% until it becomes flat. Similar to the $C_3/C_1$ ratio, this ratio stays below the Skellam limit.

Figures \ref{fig:bw_vs_omega} and \ref{fig:bw_vs_sigmaeta} also display results from the STAR Collaboration for data recorded for central collisions, 0-5\% centrality \cite{Adamczyk:2014fia,Adam:2020unf,Adamczyk:2017wsl}. 
For net proton fluctuations, the blast-wave model results are similar to STAR measurements for all three ratios. It is difficult to conclude the meaning of the solid agreement shown in the $C_1/C_2$ ratios. This agreement covers net charge, net kaons and net protons, as long as the sub-volume $\Omega$ is of the order of 50 fm$^3$ or greater. Interpreting the meaning of this agreement is difficult because it is mainly driven by having the model correctly match the particle yields with multiplicity. Unlike the $C_3/C_1$ and $C_4/C_2$ ratios there is no simple baseline for a Skellam distribution. The Skellam baseline would depend on knowing the charged particle multiplicity distributions, i.e. the moments of the multiplicity distributions for  protons plus antiprotons, kaons plus antikoans, or all charged particles.

For the net-proton distribution, the behavior of $C_3/C_1$ as a function of beam energy seems consistent with the experimental uncertainties, but the statistical errors in the experimental data are so large, that little can be concluded aside from the fact that the models predict that $C_3/C_1$  should be in the neighborhood of 0.75. These ratios from the models can show a modest sensitivity to beam energy, but any such trends are overwhelmed by the statistical errors of the experimental results at the moment.

For both net kaons and for net charge, $C_3/C_1$ and $C_4/C_2$ lie above the range of model predictions, but the large experimental error bars forbid one from stating this with great confidence. Should the experimental results with improved statistics confirm this discrepancy, it will be difficult to explain unless a significant number of charges are emitted from large clusters. In contrast, the model should always give ratios below unity for $C_4/C_2$, regardless of the choice of parameters. These failures of the blast-wave model for net kaons and for net charge are discussed in the upcoming summary, Sec. \ref{sec:summary}.

% !TEX root =  CCmoments.tex

\section{Summary}\label{sec:summary}
 
The principal goals of this study were to clarify background contributions for higher moments of charge distributions measured by the STAR Collaboration at RHIC, and to state the degree to which current experimental results are either consistent or inconsistent with these contributions. By background, this refers to sources of fluctuations besides those that arise from baryon number or charge clustering due to processes such as phase separation. The list of such sources includes charge conservation, Bose corrections, volume fluctuations, and the decays of resonances. In order to gain better insight both simple semi-analytic models with a single type of conserved charge, similar to the work performed in \cite{Savchuk:2019xfg}, and a more realistic blast-wave model which includes a more realistic accounting of the STAR acceptance, similar to what was applied in \cite{Oliinychenko:2020cmr}, were investigated. By using a highly efficient algorithm for Monte Carlo generation of particles according to the canonical ensemble, results were produced with small statistical uncertainties. This enabled the exploration of sensitivities to critical parameters of the model.

Of the various background correlations, charge conservation provided the strongest non-Poissonian contributions. Because fourth-order cumulants were defined to subtract contributions from second-order correlations, one might have expected a small contribution. Consistent with the result from \cite{Savchuk:2019xfg} for a single type of charge, it was found that generating sets of particles from a sub-volume equilibrated according to the canonical ensemble produced values of $C_3/C_1$ and $C_4/C_2$ which were significantly lower than the Skellam value of unity, which is what one would expect for uncorrelated emission. In contrast, the contribution from two-particle decays does not change either $C_3/C_1$ or $C_4/C_2$. This conclusion persisted for the more realistic blast-wave model which incorporated the conservation of all three charges and included a filter of the STAR acceptance. The correlation varied modestly according to the size of the canonical sub-volume for small sub-volumes due to canonical suppression. A modest sensitivity was also found to the spatial extent of this volume along the beam axis, as the overlay of collective longitudinal flow onto the finite acceptance in rapidity effectively lowers the probability for two balancing charges to both be observed. Due to the fact that baryon density falls with increasing beam energy, the strength of such correlations do depend on beam energy. But this sensitivity was not dramatic. Thus, even though this background contribution is rather large, it is quite smooth with respect to beam energy, so if sharp non-monotonic structures are observed experimentally with respect to beam energy, such structures are not likely to be driven by charge conservation. 

A second source of background arises from multi-particle symmetrization of the outgoing pions. By extending the recursive techniques applied for the canonical ensemble to include symmetrization, it was found that such effects should not affect the skewness or kurtosis unless the pion phase space density were to become surprisingly large. In order for the symmetrization effects to become large, the pion phase space density in the absence of symmetrization would have to double. If this were the case, the pion spectra would be more dramatically altered by Bose effects and the measured HBT radii would have to be significantly altered. 

Volume fluctuations are somewhat of a wildcard for background processes. As shown in Sec. \ref{sec:volumefluc}, such fluctuations can significantly increase the $C_4/C_2$ ratio. The STAR Collaboration invested great effort in minimizing their impact, but it is difficult to gauge the degree to which such effects might have persisted. Volume fluctuations also increase similar moments of the multiplicity distribution, which is constructed by counting charged particles rather than net charge. It is critical for the experiments to simultaneously present moments of the multiplicity distribution alongside those of the net-charge distributions. This sensitivity was also illustrated with the simple results from a system with one charge and uniform acceptance, restated in Eq. (\ref{eq:savchuk}) from the previous work of \cite{Savchuk:2019xfg}. Because volume fluctuations should similarly increase the $C_4/C_2$ ratio for net charge, net strangeness and net baryon number, behavior of such ratios for one type of charge that are not seen in other types of charge can be considered as originating from some other effect. 

Due to the way in which cumulants are constructed, the third and fourth-order cumulants should be impervious to the effects of two-particle decays. However, decays of clusters that produce four or more charged particles do contribute to $C_4$. Here, it was found that if a significant fraction of charged particles come from such decays, the higher moments can be profoundly altered, as was already known from the work in \cite{Bzdak:2018uhv}. However, the decays of such clusters would also affect the multiplicity distribution, which again underscores the importance of the experiments to simultaneously analyze fluctuation of net charge and of the the multiplicity distribution. If such effects were important, it would suggest novel contributions to the dynamics of charge production, outside of the usual paradigm of creating equilibrated distributions of hadrons. 

The results of the blast-wave calculations were displayed alongside STAR results in Sec. \ref{sec:blast}. The experimental results had much larger statistical errors than the calculations, which limits the conclusions that can be drawn. The fluctuations of net protons were not far from the range of those calculated here. This is consistent with charge conservation being the dominant source of non-Poissonian behavior, i.e. $C_4/C_2\ne 1$ and $C_3/C_1\ne 1$. By no means does this suggest that this is evidence for a lack of more novel sources of correlation in the baryons, such as that arising from phase separation. The experimental uncertainties may be currently too large to unmask such phenomena. 

The observed moments of the net-charge distribution are perplexing. Although there are large statistical uncertainties in the experimental data, it appears that $C_4/C_2$ exceeds expectations of the blast-wave calculations, and even lies above unity. Given that the net-proton distributions are roughly in line with the model predictions, this discrepancy cannot be explained by volume fluctuations, or equivalently by the systematics of event binning. Because phase transition phenomena are expected to manifest themselves mainly in the net-proton distributions, this would suggest that the decays of larger clusters into four or more charged particles might be present. If such processes were present, it would motivate a rethinking of models of chemical evolution and charge production in heavy-ion collisions. In order to confirm or dismiss this hypothesis, it is imperative that a simultaneous analysis of charge (not net charge) multiplicity distributions be undertaken from the same data sets with the same cuts on centrality. Given the greatly improved data sets currently being analyzed by STAR in the Beam Energy Scan II program at RHIC \cite{Yang:2017llt}, this puzzle should be clarified in the next year or two.

Finally, the studies here help point the way to future improvements in modeling. The picture of independent canonical sub-volumes is crude. It does incorporate the truth that charge conservation is enforced locally, over some length scale, and is sufficient to provide the understanding of how large such effects might be. However, in reality baryon number, electric charge and strangeness are created and evolve in different ways. Strangeness tends to be created early in the collision, thus allowing the balancing strange and anti-strange quarks to separate before the emission of the hadrons to which they are asymptotically assigned. In contrast, up and down quarks are more likely to be produced later in the reaction. Thus, the characteristic canonical volumes should have different sizes and different longitudinal extents depending on whether one is considering up, down or strange quarks. This is also true for off-diagonal correlations, e.g. correlations between strange and up. These only appear in the hadronic phase. A diagrammatic formalism has been developed for evolving such two-, three-, and four-body correlations as a function of the positions of each charge. These equations are based on knowing the chemical evolution of the system and the diffusion constants for each type of charge. Such calculations have been performed for two-body correlations and compared to experimentally measured charge-balance functions \cite{Pratt:2018ebf,Pratt:2019pnd} or fluctuations \cite{Aziz:2004qu}. Unfortunately, the method for three- and four-body correlations  is challenging to implement numerically \cite{Pratt:2019fbj}. It is tractable, but would require significant effort. The study presented here suggests that such calculations would be warranted if one wanted to reproduce the moments of these distributions for each type of charge, and especially if one wants to consider cross terms \cite{Abdelwahab:2014yha}, such as moments involving powers of both charge and net baryon number. Otherwise, given that charge conservation effects are expected to evolve smoothly with beam energy, one could simply see whether measurements of the ongoing Beam Energy Scan at RHIC unveil any sharp features, and assign such features to more novel types of physics. % !TEX root =  CCmoments.tex

\begin{acknowledgments}
This work was supported by the Department of Energy Office of Science through grant number DE-FG02-03ER41259 and through grant number DE-FG02-87ER40328. R. Steinhorst was additionally supported by the MSU Professorial Assistantship program and by the Director's Research Scholars program at the National Superconducting Cyclotron Laboratory. The work benefited by discussions within the BEST Collaboration, which is a topical collaboration funded by the Department of Energy Office of Science, and especially from discussions with Dmytro Oliinychenko.
\end{acknowledgments}

\end{document}